\documentclass[11pt]{article} 

\usepackage[dvips]{epsfig}

\usepackage{amsmath,amssymb}
\usepackage{amsthm}

\usepackage[top=0.8in, bottom=0.8in, left=1.2in, right=1.2in]{geometry}

\newtheorem{lemma}{Lemma}

\begin{document}



\title
{Discrete charges on a two dimensional conductor } 

\author{
Marko Kleine Berkenbusch\footnote{mkb@uchicago.edu}\\
Isabelle Claus\footnote{isabelle@control.uchicago.edu}\\
Catherine Dunn\\
Leo P. Kadanoff\footnote{leop@uchicago.edu}\\
Maciej Nicewicz\footnote{nicewicz@uchicago.edu}\\
Shankar C. Venkataramani\footnote{scvenkat@control.uchicago.edu}\\
\\
\parbox{25pc}{\small
\centering
\mbox{The Materials Research Science and Engineering Center}\\
The University of Chicago\\
5640 S. Ellis Avenue\\
Chicago IL 60637, USA\\
}
\vspace{0.5cm}
}
\maketitle

\begin{abstract}

We investigate the electrostatic equilibria of $N$ discrete charges of
size $1/N$ on a two dimensional conductor (domain). We study the
distribution of the charges on symmetric domains including the
ellipse, the hypotrochoid and various regular polygons, with an
emphasis on understanding the distributions of the charges, as the
shape of the underlying conductor becomes singular. We find that there
are two regimes of behavior, a symmetric regime for smooth conductors,
and a symmetry broken regime for ``singular'' domains.

For smooth conductors, the locations of the charges can be determined,
to within $O(\sqrt{\log N}/N^2)$ by an integral equation due to
Pommerenke [{\em Math. Ann.}, {\bf 179}:212--218, (1969)]. We present
a derivation of a related (but different) integral equation, which has
the same solutions. We also solve the equation to obtain (asymptotic)
solutions which show universal behavior in the distribution of the
charges in conductors with somewhat smooth cusps.

Conductors with sharp cusps and singularities show qualitatively
different behavior, where the symmetry of the problem is broken, and
the distribution of the discrete charges does not respect the symmetry
of the underlying domain. We investigate the symmetry breaking both
theoretically, and numerically, and find good agreement between our
theory and the numerics. We also find that the universality in the
distribution of the charges near the cusps {\em persists} in the
symmetry broken regime, although this distribution is very different
from the one given by the integral equation.

{\flushleft \emph{keywords:} Fekete points; singular domains; symmetry breaking; electrostatic equilibrium.}
\end{abstract}

\section{Introduction}
The placement of charges on a conductor is a classical problem in both
physics and mathematics.  In physics the history of the problem goes
back to the work of J.J.~Thomson \cite{TH04,BN00,BN02}. The placement
of charges on a line was investigated by Stieltjes \cite{stieltjes} in
the late 19th century. Fekete \cite{fekete} recognized the connection
between the placement of charges on a 2 dimensional domain, and
questions involving polynomial functions of a complex variable. This
problem was also investigated by Frostman \cite{frostman}, for its
connections to Potential theory. In general, the problem is to place
$N$ charges of strength $1/N$ on a surface so that the energy is
minimized.  In this placement the force component parallel to the
surface for each charge vanishes.

This problem has quite a different nature depending upon the dimension
of the system.  In three dimensional space, one uses an inverse square
force law.  The placement upon any surface, even one as simple as the
surface of a sphere is quite complex.  The discrete charges form an
approximation of a lattice with many defects
\cite{erber,BN00,BN02,sphere_packing}.  The solution might properly be
described as being chaotic.

In contrast, in two dimensions, one finds smoother behavior.  The problem is
one of placement of lines of charge on the curve which bounds a two-
dimensional conductor.  Forces between the charges are given by an inverse
first power of the distance.   

The analogous continuum problem has a very good
general formulation via the Riemann mapping theorem.  To calculate the 
distribution of continuum charges on the exterior of a simply connected region
in the plane, construct the unique function
$F(w)$ that takes the exterior of the unit circle into the exterior of the
region, and has the property that
$F(w)
\rightarrow w~$ as 
$~ w \rightarrow \infty$.  Then the density of charges at the point
$z=F(w)= F(e^{i\theta})$ on the surface of the conductor is given by 
\begin{equation}
   \rho(\theta)= |w F'(w)|^{-1}
\label{charge}
\end{equation}
This result provides the basis for all subsequent work on discrete charges.  

More recent work \cite{faraday_cage,fekete_review,KG74} has looked at the 
placement of discrete charges
in terms of the placement of points on the circle at
\begin{equation}
   w_j= e^{2\pi i \theta_j}
\end{equation}
The $\theta_j$'s are real.  According to a set of important recent
theorems \cite{fekete_review}, for sufficiently smooth curves $C$ , we
can determine the charge placement for all large enough values of $N$
in terms of a smooth periodic function, $\psi(\theta)$\footnote{This
functional is conventionally denoted by $\Phi$. We however reserve
$\Phi$ for the electrostatic potential}, by writing
\begin{equation}
\theta_{j} \approx \theta_0 + \frac{2 \pi j}{N} + \frac{1}{N} \left[ \psi\left(\theta_0 + \frac{2 \pi j}{N}\right) - \psi(\theta_0)\right] 
\label{place}
\end{equation} 
With appropriate choice of $\theta_0$, the expression \ref{place}
gives the placement of charges with error of order
$\sqrt{\log(N)}/N^2$ \cite{Pom69a} for all extrema of the energy
\begin{equation}
   E^N(\{\theta\})= -\frac{1}{2 N^2}\sum_{j\ne k} \Re ~ \ln[ F(w_j)-F(w_k)]
\end{equation}
Here we represent the usual logarithmic potential for line-charges
by taking the real part of the complex logarithm function.

\subsection{ New empirical results}
We calculate the
actual position of charges on two different kinds of shapes: ellipses
and hypotrochoids. These two kinds of shapes are defined as the image
of the unit circle under maps of the form
\begin{equation}
F(w)=w-\frac{c}{w^r}
\label{map}
\end{equation}
Here $r=1$ for ellipses and $r=2$ for hypotrochoids.  The parameter $c$
controls how different these figures are from the circle.

\begin{figure}[htbp]
\includegraphics[width=9.5cm]{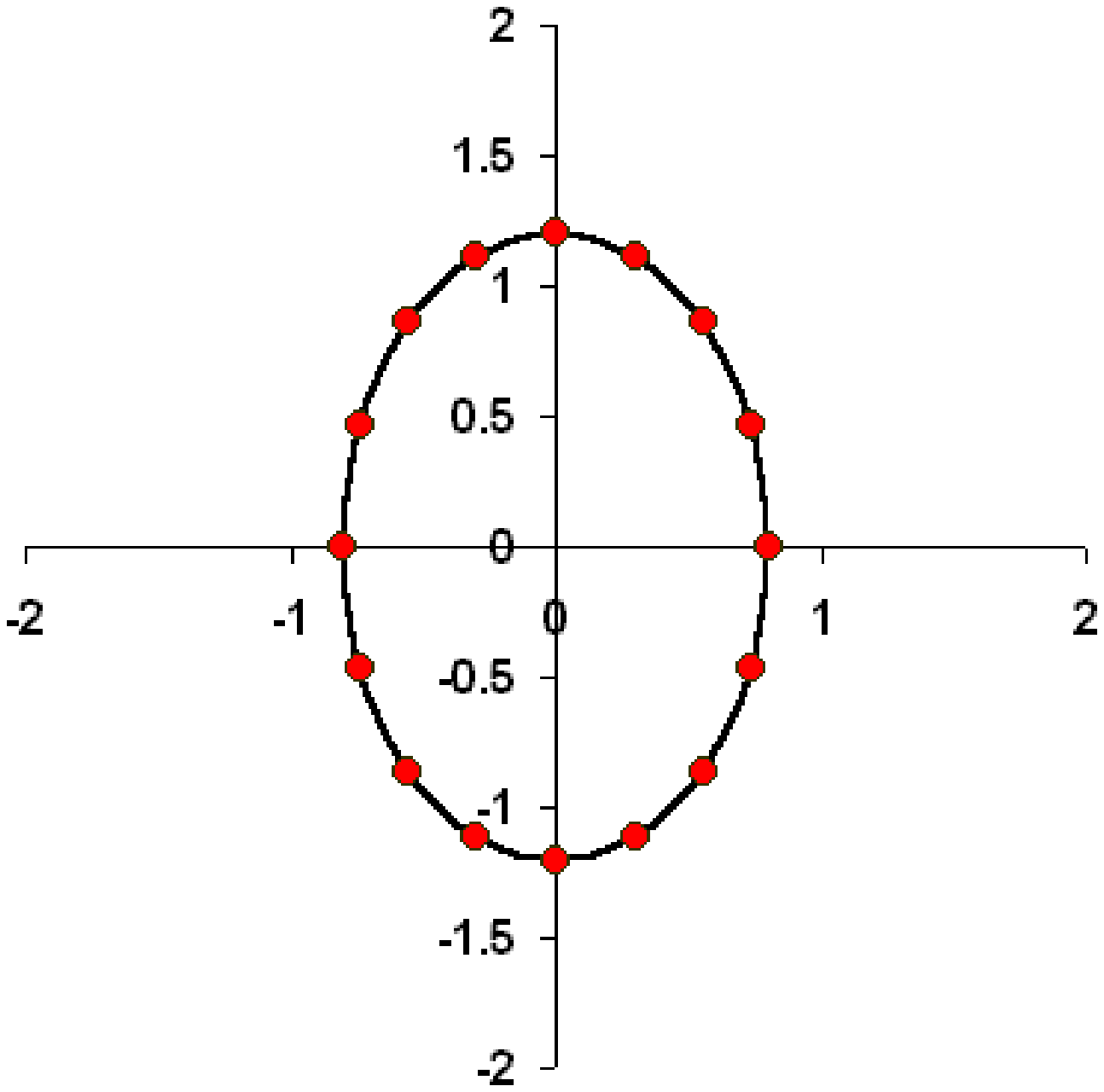}
\hfill
\includegraphics[width=9.5cm]{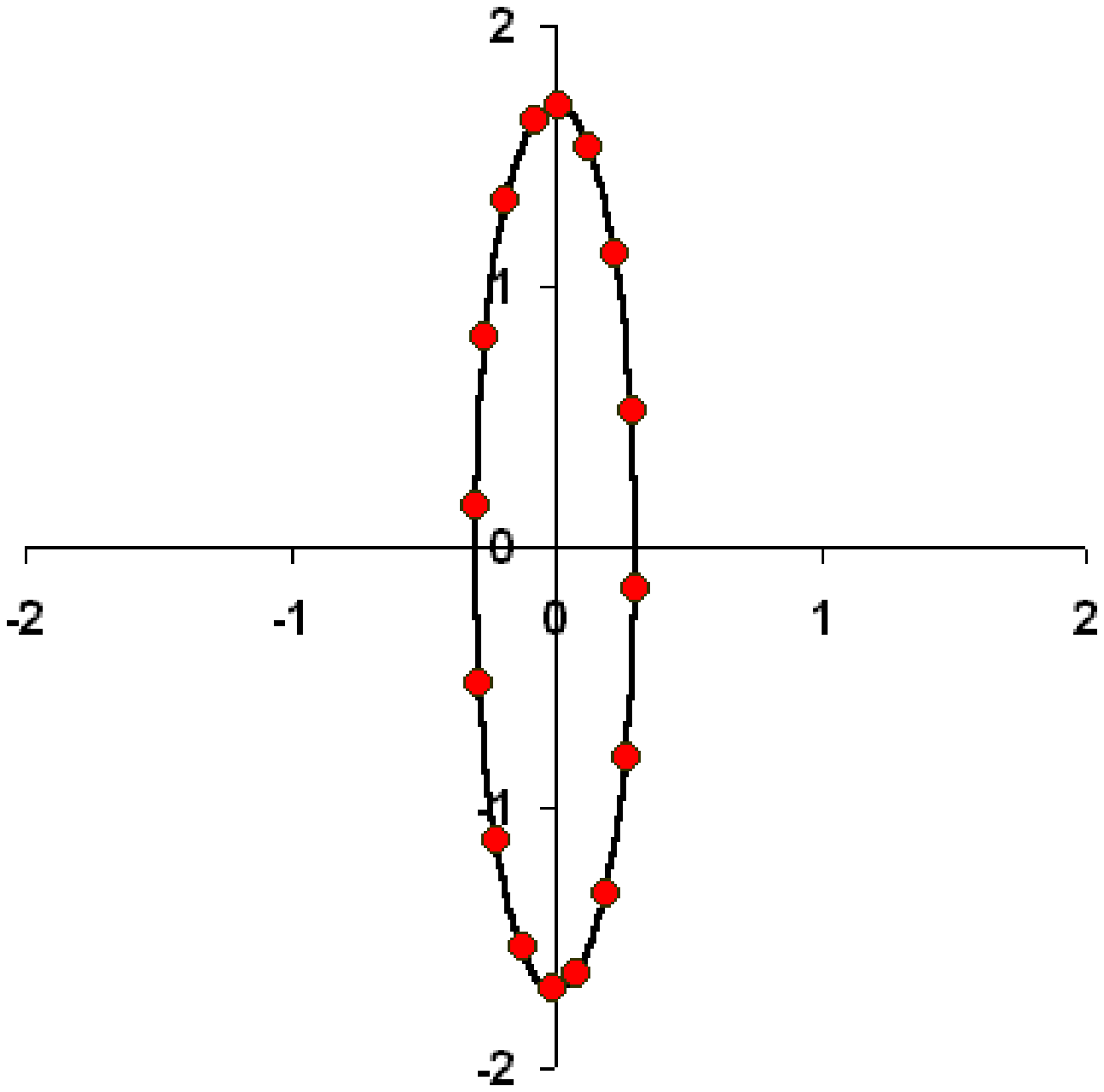}
\caption{ Placement of 16 charges on two different  ellipses, 
$c=0.2$ and $c=0.7$ respectively.  Notice the
breaking of the parity symmetry in parts b. The ellipse degenerates to a line 
when $c=1$.} 
\label{ellipse1}
\end{figure}    

\begin{figure}[htbp]
\centering
\includegraphics[width=6cm]{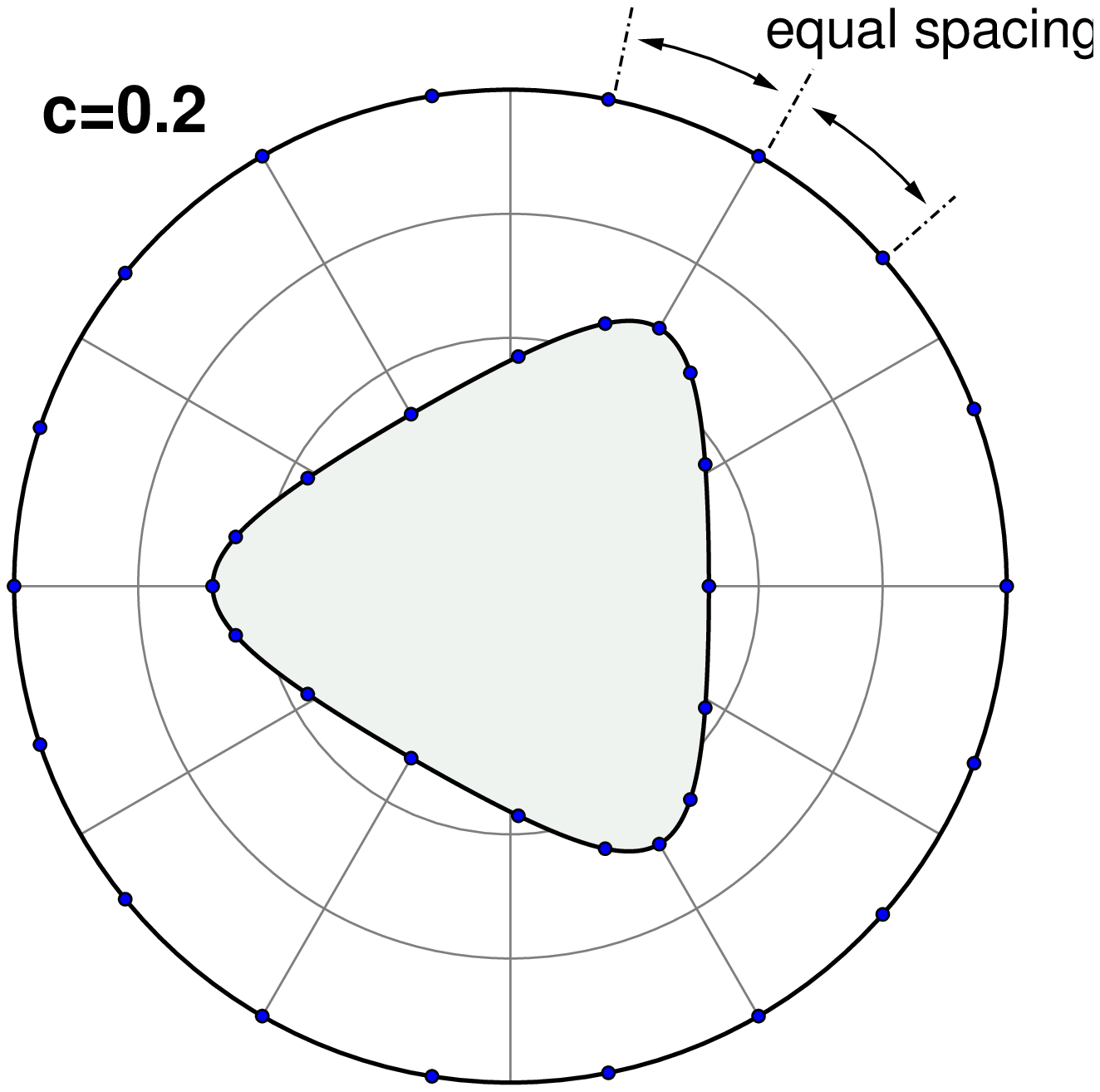}
\hfill
\includegraphics[width=6cm]{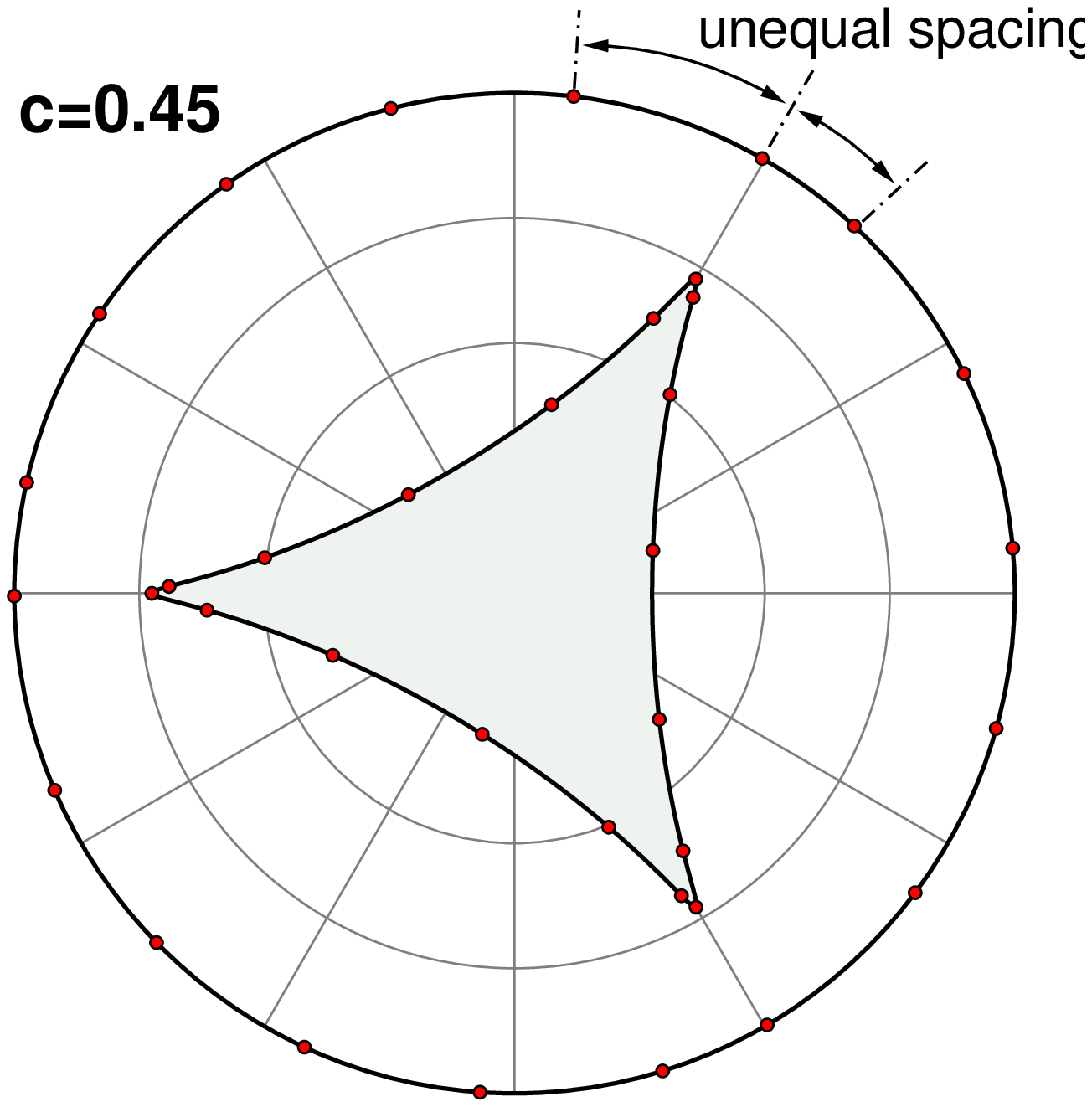}\\
\includegraphics[width=6cm]{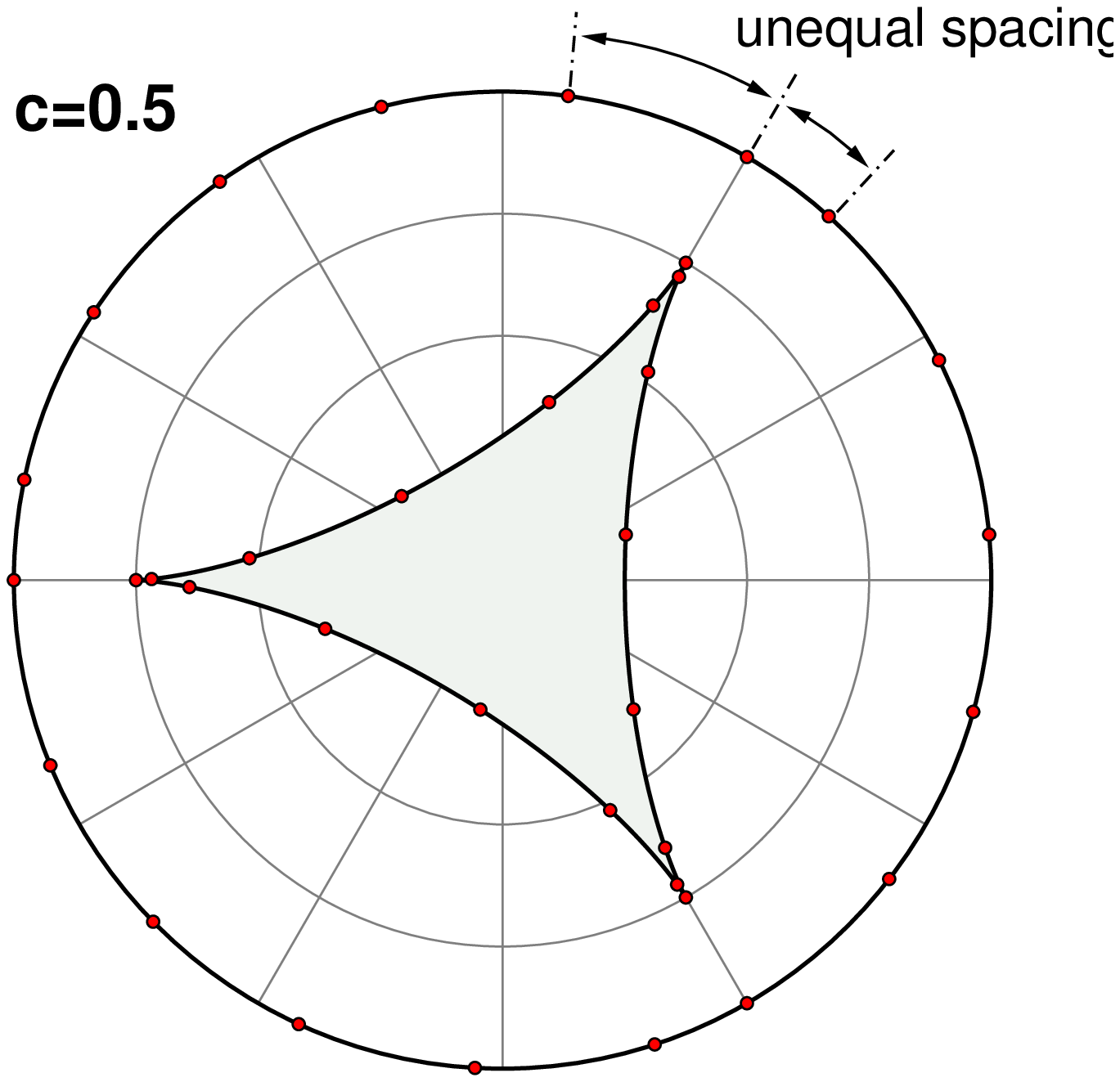}
\caption{ Placement of 18 charges on three different hypotrochoids, 
corresponding to $c=0.2$, $c=0.45$ and $c=0.5$ respectively.  Notice
the breaking of the parity symmetry in part b and c. The hypotrochoid is 
smooth for $0 \le c < 1/2$ but gains three pointed cusps at $c=1/2$.} 
\label{hypo1}
\end{figure}

Results of
these mappings are shown in figures 
\ref{ellipse1},\ref{hypo1}. These figures also include the positions of the
charges placed upon these conductors.  As one might expect from equation 
\ref{charge}, the charge
densities are highest in the high curvature areas of the bounding curves. 
However, the figures also
contain some surprises.  In both cases, there is a symmetry breaking in
the more pointed figures. In this breaking, the left-right symmetry is
broken at each pointy region, and the overall parity symmetry of the
figure is lost. We shall explore this symmetry-breaking in more detail
below.

\begin{figure}[htbp]
\centering 
\includegraphics[width=8cm]{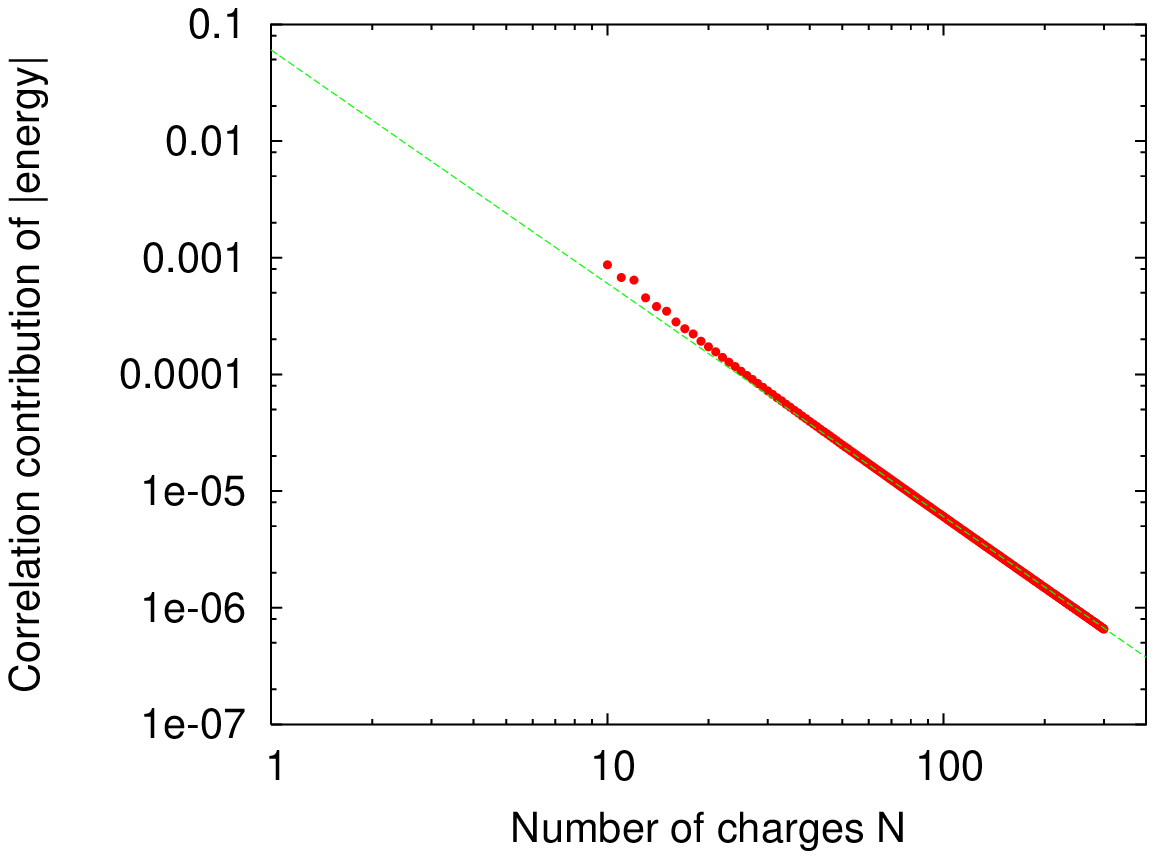}\\
\includegraphics[width=8cm]{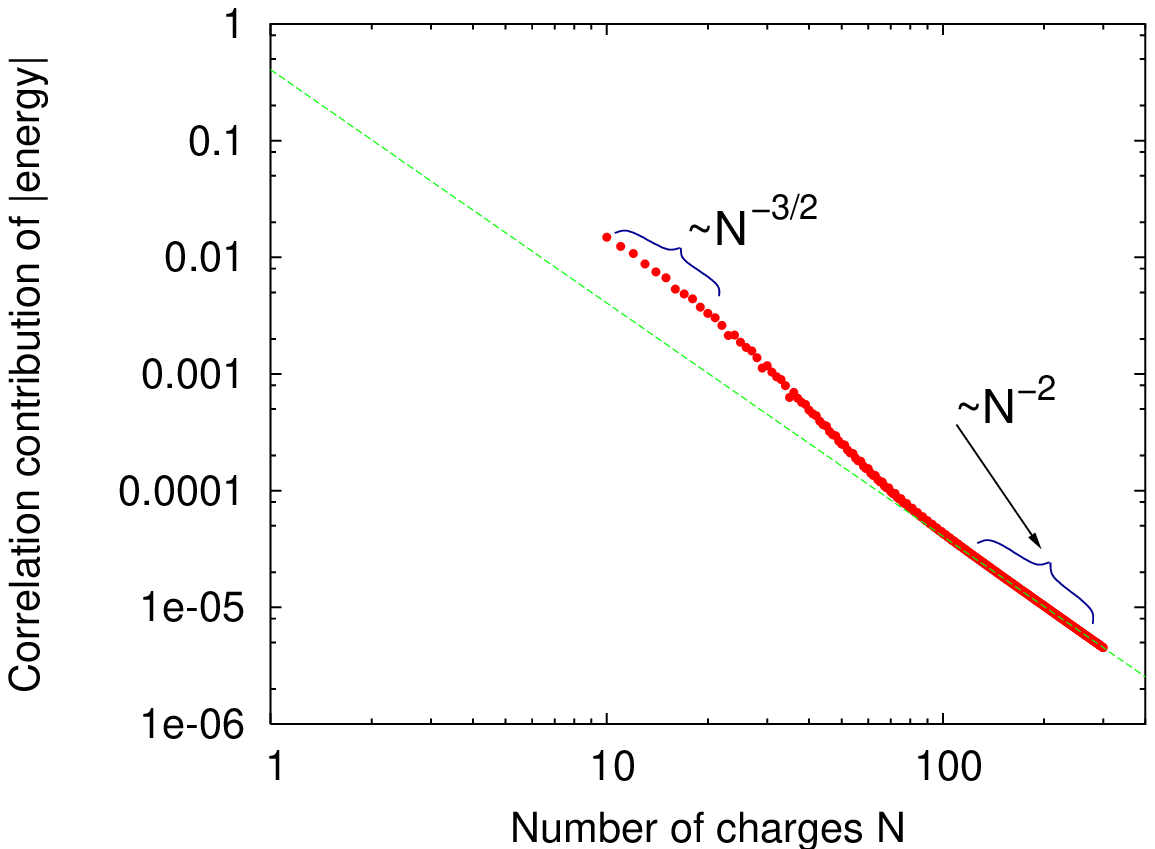}\\
\includegraphics[width=8cm]{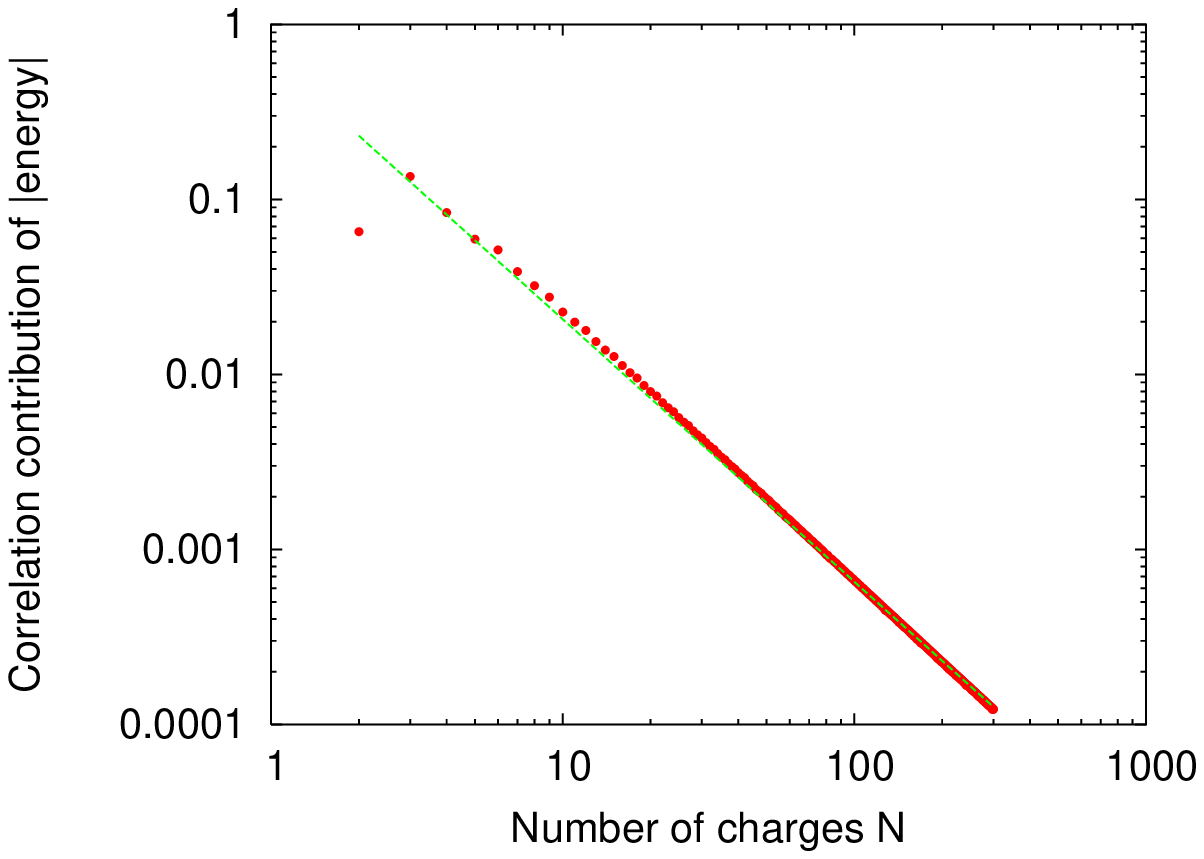}
\caption{ Plot of correlation energy versus $N$.  The three curves shown
correspond to the three hypotrochoids shown in figure \ref{hypo1}.} 
\label{Ecorr}
\end{figure}

To understand these charge placements, look at the  energies they
generate.  When the sizes of the figures are properly adjusted the
continuum correlation energy is zero. The next term in the energy is the
self-energy:
\begin{equation}
E_s^N = - \frac{\ln N}{2 N}
\end{equation}
This term is independent of charge placement.  The remaining energy is
a correlation energy, and has a power-law behavior in $N$ for large $N$:
\begin{equation}
E_c^N  \sim  N^{-s}
\label{E_c}
\end{equation}
Here the exponent $s$ is known to be two for smooth curves. In
figure \ref{Ecorr} we plot the correlation energy versus $N$ for  three
different hypotrochoids.  In the first case, we can clearly that $s=2$. 
For the singular figure, corresponding to figure \ref{hypo1}c,  the
log-log plot exhibits a slope of 1.5.  In the intermediate case,
corresponding to the blunted points of figure \ref{hypo1}b, there is a
crossover from  $s=1.5$ for smaller $N$ to $s=2$ for larger $N$.

\subsection{ Connection to Modern work in dynamical systems theory}
Note the close connection between equation \ref{place} and the
expression giving the positions of points in long cycles and
quasi-periodic orbits of KAM \cite{Arnold} theory.  KAM theory
provides an expansion about continuum motion, in which the positioning
of points in a cycle closely follows that of the corresponding
continuum orbit. In this Coulomb system too, a near continuum result
gives a charge density which is only a little different from the one
in the continuum case.  As in KAM theory, equation \ref{place}
describes a near-continuum result by giving a function, here $\psi$,
which describes the deviation from and permits a smooth passage to the
continuum limit in which $N$ goes to infinity.

In KAM theory, the cycles closely follow a continuum orbit.  So the
continuum theory predicts the curve on which the points will occur;
but it does not give the positioning of the points in the orbit.  A
more subtle calculation is required to find that.  Here the analogous
fact is that the initial angle $\theta_0$ 
of equation \ref{place} is not determined by any continuum theory.  In fact, one
does not know a first principles method of finding $\theta_0$.

Dynamical systems theory is much concerned with bifurcations, {\em
i.e.}, situations in which a qualitative change in behavior occurs.
In the Coulomb problem there is a bifurcation from the single {\em
symmetric} to multiple solutions and symmetry breaking behavior.  In
dynamical systems theory when a parameter is varied, bifurcations can
occur for longer and longer orbits producing an interesting kind of
critical behavior \cite{Gree79,Kad81,Kad82}. Here the analogous
situation arises when the bounding curve develops cusps as the
parameter $c$ is varied, and close to this cusp formation, the Coulomb
system exhibits a symmetry breaking bifurcation. This behavior
produces interesting anomalies including the cross-over in the
$N$-dependence of the energy.

\subsection{ Paper Outline}
Section~\ref{sec:real_space} of this paper deals with the
equations for the placement of charges on the circle and on the circle
mapping onto "real space".  
We find that, for smooth curves, the uniform placement
on the circle gives a correlation energy which goes exponentially to
zero for large-$N$.
We derive an integral equation which describes the ``continuum''
behavior of the deviation of the charges from the uniform placement on
the circle. This is related to, but different from, the equation
formulated by Pommerenke \cite{Pom69a}. Our derivation gives a
physical interpretation of the equation as being related to the
distortions of fields produced by the discrete charges' self-energy
effects. 

Section~\ref{sec:fourier} follows the analysis of the integral
equation in Fourier space. Here the first quantity of interest is the
structure function $s_k$, which is the Fourier transform of the charge
distribution.
The equation is solved in the large $N$-limit for both the ellipse (an
old result \cite{Pom67}) and the
hypotrochoid--a new result.  The solution, involving an elliptic
function, is formally very similar in the two cases. We also compare
the solutions we obtain with numerical observations, and we find that
the Integral equation is not applicable for ``singular'' shapes, and
in the symmetry broken regime.

In Sec.~\ref{sec:singular}, we present new results for the singular
limit of the Coulomb problem. 
We investigate the energy and the charge placement in the symmetry
broken regime, {\em i.e.}, for ellipses with $c \approx 1$,
hypotrochoids with $c \approx \frac{1}{2}$. We also study ``singular''
curves with cusps and corners, {\em viz.} the line segment, the
hypotrochoid with $c = \frac{1}{2}$ and various regular polygons. 
We present scaling arguments for the dependence of the energy on $N$
in the singular limit, and use this to obtain scaling laws for the
symmetry-breaking transition. We also compare these results with
numerical simulation.
\section{Real Space Analysis} \label{sec:real_space}
  
\subsection{On the circle}
We describe our electrostatics problem by using complex variables.  Thus a
charge of size $1/N$  placed at the point $(x_j,y_j)$ will generate the
complex potential at the point $(x,y)$ via the formula  
$$
\Phi(z)=-\frac{1}{N} \ln (z-z_j) 
$$
where $\ln$ is the complex logarithm and $z$ is $x+iy$.
 If there are
$N$ such charges 
\begin{equation}
\Phi(z)= - \frac{1}{N}  \sum_{j=1}^{N} \ln (z-z_j)
\label{pot}
\end{equation} 
When the $N$ charges are uniformly distributed on
a circle of radius one,  then the potential  is 
\begin{eqnarray}
\Phi(z) &=& -\frac{1}{N} \sum_{j=1}^{N} \ln (z-e^{2\pi i
j/N}) \label{fac1} \nonumber \\ &=& -\frac{1}{N}\ln  ( z^N-1) 
\label{p1}
\end{eqnarray}     

Note the elegance of the result of equation (\ref{p1}).  In the limit
of large $N$, inside the unit circle $z^N$ is very small, and the
potential is well approximated by $\Phi= (-\pi i)/N$.  Outside, we
find $ \Phi = -\ln z $ which is the continuum result. To get the
potential seen by the charge at $z=1$, we have to subtract away the
charge potential it produces and use
\begin{equation}
\Phi^0(z) = \frac{1}{N} \ln \frac {z-1}{z^N-1}
\label{phi01}
\end{equation}
Equation (\ref{phi01}) describes the potential upon our charge as the potential
produced by the charges on the circle less the potential generated by our
charge itself.   In the neighborhood of  
$z=1$, 
  we can evaluate this potential    
 as
$$\Phi^0(z)  = -\frac  {\ln N} N  - \frac {N-1}{2N}
(z-1)-\frac{(N-1)(N-5)}{24N} (z-1)^2 + \dots. 
$$

The total energy of this set of charges is obtain as one half this
potential (at $z=1$) times the charge summed over all charges. The result is
\begin{equation}
E_N= -\frac  {\log N} {2N}  \equiv E_s
\label{Eself}
\end{equation}
We previously described  this terms as a self-energy.  More precisely the
energy of our charges is the energy generated by a continuous circle of
charge, namely zero, minus the energy needed to put together the
``missing'' portions of the circle.  Those portions are the lines of charge
counted as not present in expressions like equation (\ref{phi01}).  These
are lines of charge of length $2\pi/N$ not generating potential in the gap
around our charge itself.  The energy of one such gap is $(\log N) /2 N^2$. 
Subtracting away $N$ such energies gives expression (\ref{Eself}).

\subsection{Uniform Spacing on the circle} \label{complex}
\begin{figure}[t]
\centering  
\includegraphics[width=10cm]{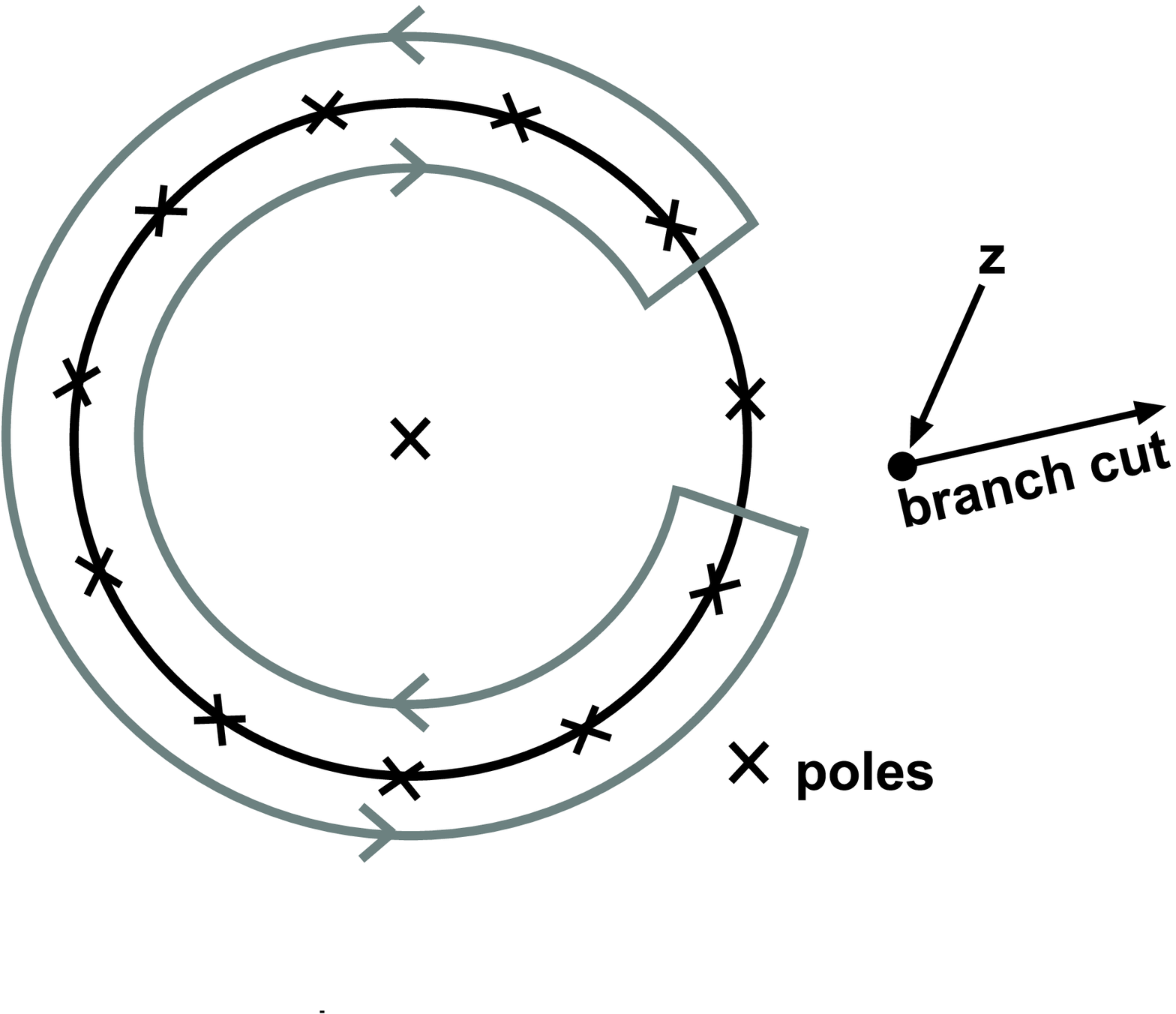}
\caption{ Contour for calculating the potential generated by $N-1$
charges on the circle.} 
\label{contour_1}
\end{figure}

This polynomial method is elegant but it is difficult to generalize to other
curves beyond the circle.  A method less elegant but more susceptible to
generalization is to evaluate the potential  via the
contour integral:
\begin{equation}
\Phi^0(z)= -\int_C \frac{dy }{2 \pi i y
 } \frac{1}{y^N-1}  \log(z-y)
\label{c1}
\end{equation}
Here the contour, $C$, encircles all the poles of $1/(y^N-1)$ except the
pole at $y=1$ (See Figure \ref{contour_1}).  By summing the
contributions from the various poles we find the result of equation
(\ref{fac1}) once more.

An alternative mode of evaluation is to deform the contour to include
the poles at zero and one and the branch line which runs from $z$ to
infinity.  The result is then
\begin{equation}
\Phi^0(z)= \log z  + \frac{\log (z-1)}{N} -\frac{1}{N} \int_{z^N}^\infty
\frac{ds }{ s} \frac{1}{s-1}  
\label{int1}
\end{equation}
which works out to the result (\ref{p1}) once more.

\begin{figure}[t]
\centering  
\includegraphics[width=10cm]{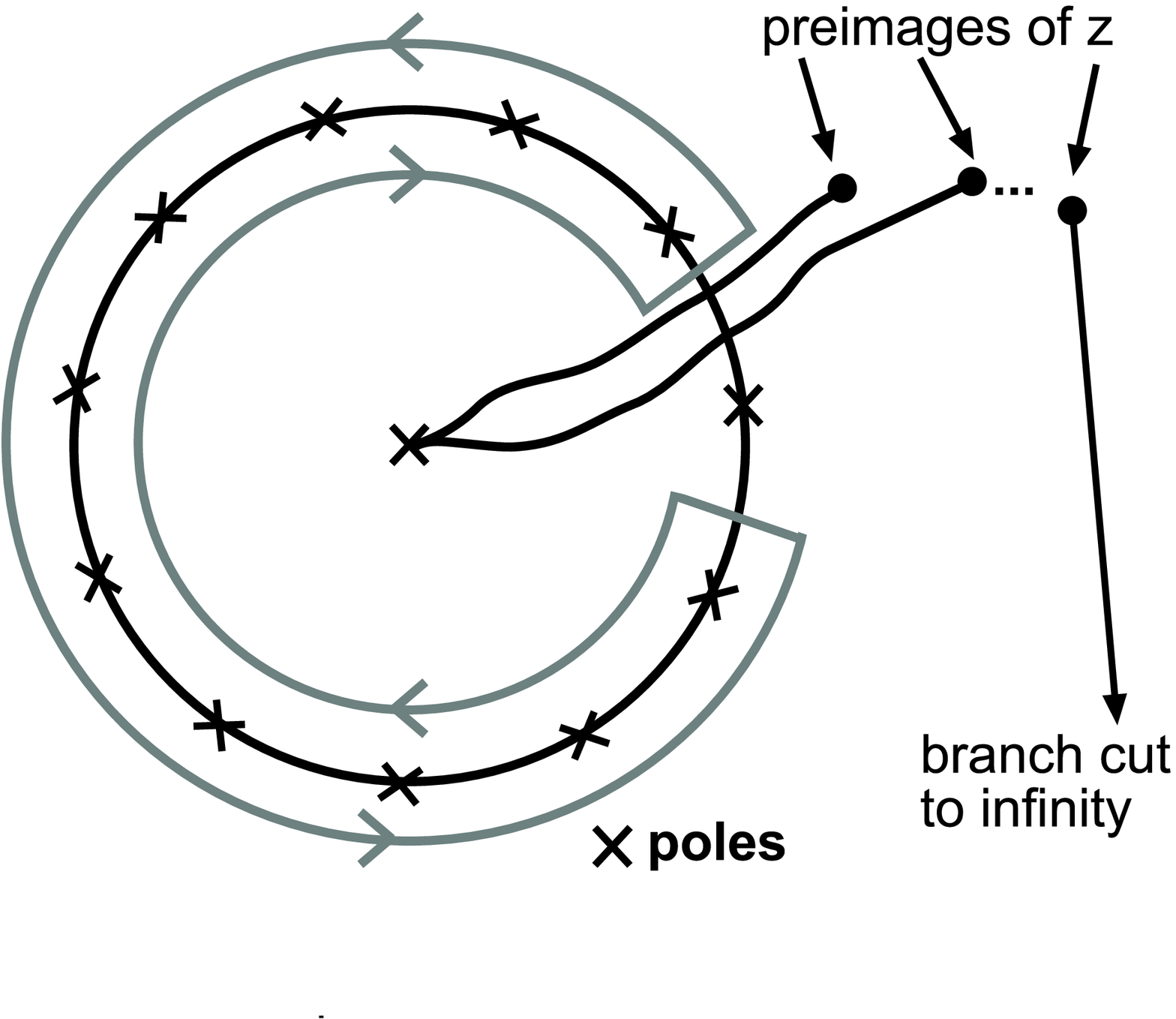}
\caption{ The contour for equation (\ref{c2}). The main difference from
the previous figure is that there are some branch lines within the unit
circle. This figure is drawn with $v=1$.} 
\label{contour_2a}
\end{figure}

The next step is to generalize 
the calculation to include the possibility of a curve other than the unit
circle. Let $F(w)$ be the function which conformally maps the exterior of
the unit circle into the exterior of a curve $C$. We shall place our charges
on the curve $C$, but give them the placement which is appropriate for the
mapping of charges uniformly spaced on the unit circle.  
        Now consider once more the effect of $N-1$ charges on $C$.  Instead of
the integral (\ref{c1}), the potential on a given charge is set by the
more general object
\begin{equation}
\Phi_N(z,v)= - \int_C \frac{dw }{2 \pi i w
 } \frac{w^N}{w^N-v^N}    \log(z-F(w))
\label{c2}
\end{equation}   
Here $v$ is on the unit circle 
and the contour, $C$, encircles the points 
\begin{equation}
w_j= v e^{2 \pi i j/N}  \quad
    \text{for}  \quad  j=1,2,\dots,N-1.
\label{w} 
\end{equation}
in the positive sense (See figure \ref{contour_2a}.)
 The real part of $\Phi_N(z,v)$ is the potential at the space point $z$,
produced by the specified set of charges. Note that this potential does not take
into account the redistribution of charges required to produce an
equilibrium distribution. 

\begin{figure}
\centering  
\includegraphics[width=10cm]{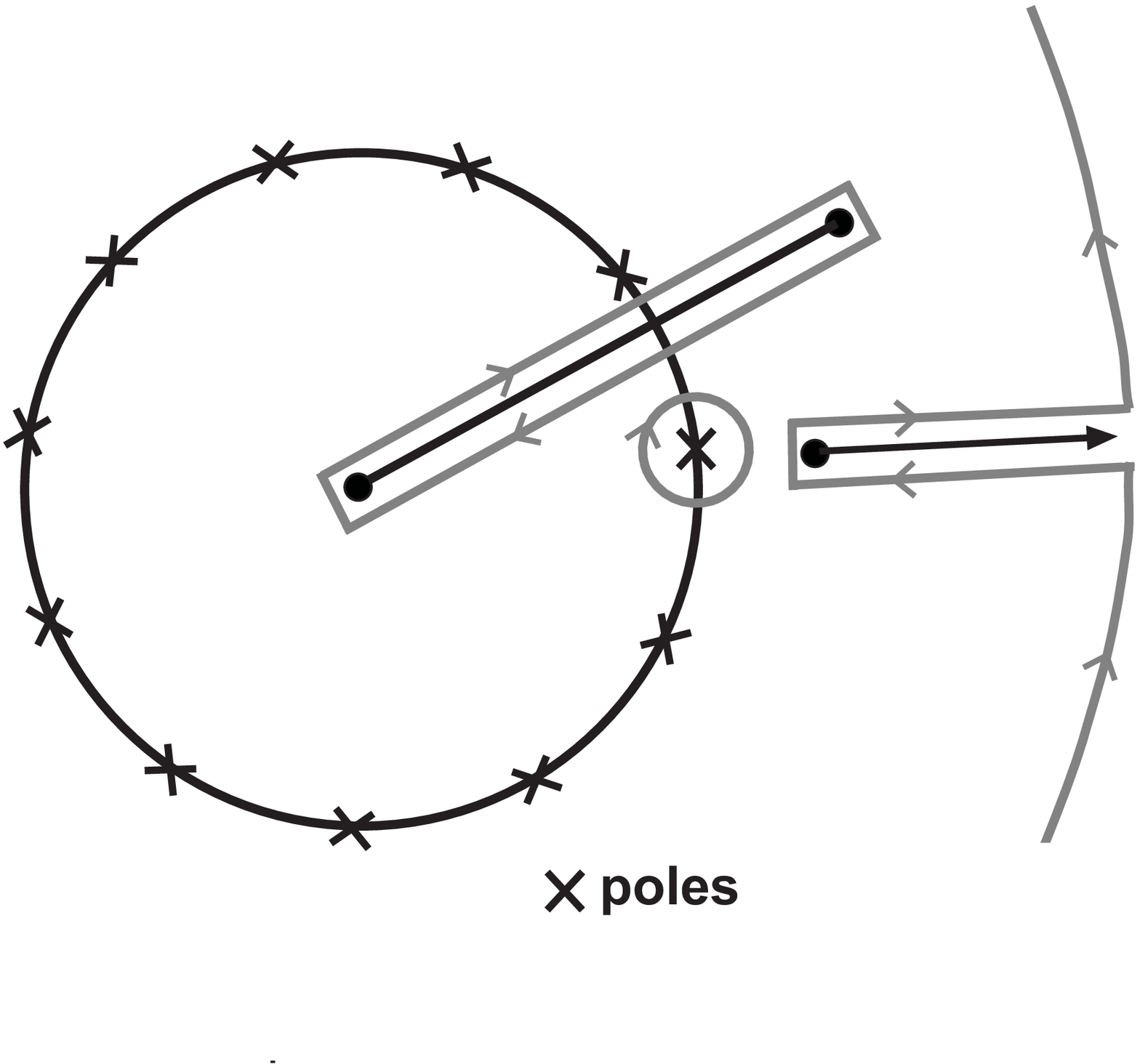}
\caption{ The contour for equation (\ref{c2}) after deformation.} 
\label{contour_2b}
\end{figure}

The calculation follows much the same lines as the previous one, but the
details of the calculation are a bit different because the integrand
is different.   As we move the contour we get four different
contributions to the result (\ref{c2}) (See Figure \ref{contour_2b}.) The first
contribution comes from the encircling of all the branch lines of the
logarithm.  The number of the branch lines that connect the preimages of z to 0 depends on the map.
The next contribution comes from encircling the pole at $w=v$. This gives a contribution from the
``missing'' charge, which is 
$$
\frac{1}{N} \log [z-F(v)]
$$
Next comes a contribution from circling the branch line which goes to the point at infinity.  Because that
encirclement contribution has a logarithmic divergence, we do
not carry it out all the way to infinity, but instead cut it off at some
large but finite length $L \gg 1$. Then we also need a contribution from a
circle at infinity.  All this is illustrated in figure \ref{contour_2b}.  The branch
line contribution is
\begin{eqnarray}
  \int_{F^{-1}(z)}^L \frac{dw }{ w
 } \frac{w^N}{w^N-v^N}  = \frac 1N  \int_{[F^{-1}(z)]^N}^{L^N}
\frac{du }{ u
 } \frac{u}{u-v^N}   \nonumber \\ 
=  {\frac {1}{N}}  ~ \log \frac {L^N-  v^N}  {[F^{-1}(z)]^N- v^N}
\nonumber \\
\approx   ~ \log  L  -{\frac {1}{N}} \log [ {[F^{-1}(z)]^N- v^N}]\nonumber
\end{eqnarray}
Finally, the large circle gives just 
$$
-\log L
$$   
since for large $w$,  $F(w)$ is approximately $w$.  When we add up
all these terms we see that
\begin{equation}
\Phi_N(z,v) = \frac{1}{N} \log(z-F(v)) - \frac{1}{N} \sum_{u\in F^{-1}\{z\}} \log (u^N - v^N)
\label{phi}
\end{equation}

Equation (\ref{phi}) is a most interesting result.  For $z$ inside the curve
$C$,  the potential is 
$$ -\log (-v) + \frac 1N \log[z-F(v)]$$
The first term is an imaginary constant describing how we placed the
branch line.  The second is a contribution of the ``missing" charge.  
For $z$ outside the the curve $C$
the potential is 
$$ -\log ~ [F^{-1}(z)] + \frac 1N \log[z-F(v)]$$ The first term
defines the potential in the limit in which $N$ goes to infinity.
This is the familiar but deep statement that the inverse of the
mapping is proportional to the exponential of the complex potential.
The second term is once more the missing charge.

Finally, we can look at the case in which the potential is evaluated at the unrelaxed position of
the $N$th charge, i.e. $z=F(v)$.  We then have:

\begin{displaymath}
\Phi_N(F(v),v) = \lim_{z\rightarrow F(v)}\Phi_N(z,v)=\lim_{z\rightarrow F(v)} \frac{1}{N} \log \left[ \frac{z-F(v)}{\prod_{u \in F^{-1}\{z\}} (u^N-v^N) }\right]
\end{displaymath}
Clearly, it is $v \in F^{-1}\{F(v)\}$, so all but one part of the limit can be trivially evaluated.  The one piece in which $ u = v$ can be done by l'Hospital's rule.  It then follows that:

\begin{equation}
\label{eqn:phin}
\Phi_N(F(v),v) = \frac{1}{N} \log \left[ \frac {F^{\prime}(v)}{Nv^{N-1}} \right] - \frac{1}{N} \sum_{u \in F^{-1}\{F(v)\}, u \neq v} \log ( u^N - v^N )
\end{equation}

Now we can write down the total energy of the system.  Since

\begin{displaymath}
E_N = \frac{1}{2N} \sum_{j=1}^N \Phi_N(F(v_j), v_j)
\end{displaymath}

we have

\begin{displaymath}
E_N = -\frac{\log N}{N}+\frac{1}{2N^2}\sum_j\log \left[\frac{F^{\prime}(v_j)}{v_j^{N-1}}\right]-\frac{1}{2N^2}\sum_{j=1}^N\sum_{u \in F^{-1}\{F(v)\}, u \neq v} \log (u^N-v^N)
\end{displaymath}

In general, it is difficult to find all such $u's$ and perform the sum.  However, in the case of the ellipse given by $F(z) = z + \frac{c}{z}$, this expression nicely simplifies.

\subsection{Ellipse energy - uniform spacing}
To begin, let us evaluate the first sum in the above expression for the uniform spacing of charges, i.e. put
$v_j = \zeta_j$, where $\zeta_j$'s are the $N^{th}$ roots of unity.  

We have,

\begin{displaymath}
\frac{1}{2N^2}\sum_{j=1}^{N}\log(\zeta_jF^{\prime}(\zeta_j)) = \frac{1}{2N^2}\log \prod_{j=1}^N ( \zeta_j-\frac{c}{\zeta_j}) 
\end{displaymath}

It follows that

\begin{displaymath}
\prod_{j=1}^N ( \zeta_j-\frac{c}{\zeta_j}) = (-1)^N \prod_{j=1}^N (1-\frac{c}{\zeta_j^2})
\end{displaymath}

We notice that since the roots of unity form the cyclic group of order N, and taking N to be odd, 2 always has a multiplicative inverse in $\mathbf{Z}_N$, 
and the group is invariant under squaring the roots of unity.  

Therefore, for odd N,  
\begin{displaymath}
(-1)^N \prod_{j=1}^N (1-\frac{c}{\zeta_j^2}) = -\prod_{j=1}^N (1-\frac{c}{\zeta_j}) = -\prod_{j=1}^N (1-c\zeta_{N-j}) = 
-c^N \prod_{j=1}^N (\frac{1}{c}-\zeta_{N-j}) = 
\end{displaymath}

\begin{displaymath}
 = -(1 - c^N)
\end{displaymath}

It follows
\begin{displaymath}
\frac{1}{2N^2}\sum_{j=1}^{N}\log(\zeta_jF^{\prime}(\zeta_j)) = \frac{1}{2N^2}\log(1-c^N)
\end{displaymath}

If N is even, then $2 * \mathbf{Z}_N \cong \mathbf{Z}_{N/2}$ so that
\begin{displaymath}
(-1)^N \prod_{j=1}^N (1-\frac{c}{\zeta_j^2}) = \prod_{j=2,4,6...}^{N}(1-\frac{c}{\zeta_j})^2 = (1-c^{N/2})^2
\end{displaymath}

and therefore
\begin{displaymath}
\frac{1}{2N^2}\sum_{j=1}^{N}\log(\zeta_jF^{\prime}(\zeta_j)) = \frac{1}{N^2}\log(1-c^{N/2})
\end{displaymath}

Please note that the evaluation of this sum can be readily generalized to the family of curves given by $F(z) = z+\frac{c}{z^{(p-1)}}$ where p is prime.  Then $zF^\prime(z) = z-\frac{(p-1)c}{z^{(p-1)}}$, and of course, as before, if $N \neq 0$ $mod$ $p$, raising the $N^{th}$ roots of unity to the $p^{th}$ power simply reshuffles them so that

\begin{displaymath}
\frac{1}{2N^2}\sum_{j=1}^{N}\log(\zeta_jF^{\prime}(\zeta_j)) = \frac{1}{2N^2}\log(1-((p-1)c)^N)
\end{displaymath} 

In the case that $N = 0$ $mod$ $p$,  

\begin{displaymath}
\prod_{j=1}^N (1-\frac{(p-1)c}{\zeta_j^p}) = \prod_{j=p,2p,3p...}^N (1-\frac{(p-1)c}{\zeta_j})^p = (1-((p-1)c)^\frac{N}{p})^p
\end{displaymath} 

and we conclude that 

\begin{displaymath}
\frac{1}{2N^2}\sum_{j=1}^{N}\log(\zeta_jF^{\prime}(\zeta_j)) = \frac{p}{2N^2} \log (1-((p-1)c)^\frac{N}{p}).
\end{displaymath} 

Now we just need to evaluate the second sum for the ellipse and we will have the exact energy expression for a uniform arrangement of charges.  
To this end we need to obtain the $u's$ which satisfy the equation $F(\zeta_j) = F(u)$ or $u^2 - uF(\zeta_j) + c = 0$.  It is clear that $u = \zeta_j$ satisfies the equation, and from this it follows that, since the product of the two roots of the above quadratic is c, the other $u$ must be $\frac{c}{\zeta_j}$.

We now have:
\begin{displaymath}
\frac{1}{2N^2}\sum_{j=1}^N\sum_{u \in F^{-1}\{F(v)\}, u \neq v} \log (u^N-v^N) = \frac{1}{2N} \log(1-c^N)
\end{displaymath}

since $\zeta_j^N$ = 1.

Finally, for odd N,

$$E_N = -\frac{\log N}{N} - \frac{1}{2N}\log(1-c^N) + \frac{1}{2N^2}\log (1-c^N)$$

while for even N, we have:

$$E_N = -\frac{\log N}{N} - \frac{1}{2N}\log(1-c^N) + \frac{1}{N^2}\log (1-c^\frac{N}{2})$$

The remaining material in this chapter should be considered to be more
heuristic in content.

\subsection{Heuristic Discussion of Energies}

We have just completed a careful analysis of the self-energy in the
case in which the charges are uniformly distributed on the unit
circle. Now we extend this discussion to include the effects of a
non-constant charge density on the circle.  In fact, if we define the
charge density to be the smooth function of $j$ defined so that
\begin{equation}
\theta_J-\theta_K = \int_K^J  \frac{dj}{N \rho_j} 
\label{density}
\end{equation} 
If $\rho_j$ is independent of $j$ we know that equation (\ref{Eself}) gives an
essentially exact answer for the large-$N$ case.

Our result for contributions of the self-energy to the potential for a
single particle is given by equation (\ref{eqn:phin}), which can be written in terms
of the  spacing between particles on the curve $C$ which is given by a spacing
factor on the circle times the conversion factor from circle it curve $1/F'$. 
Thus the spacing is
\begin{equation} 
\delta=\frac{|vF'(v)|}{ 2 N \pi\rho }   
 \label{space}
\end{equation}
If we take our original expression for the self-energy contribution to
the potential, and then say that this contribution to the real potential
depends only on the actual spacing we find a potential contribution    
\begin{equation}
 \phi_s(j) =\frac{1}{N}  \log \left[\frac{|F'(v_j)|}{2 N \pi \rho_j} \right]  
\label{pot_s}
\end{equation}
Just as before, we calculate a contribution to the total energy by taking
half the potential  multiplied by the charge density and adding.  We find
\begin{equation}
 E_s = \frac{1}{2N} \sum_{j=1}^N  
 \log \left[\frac{|F'(v_j)|}{2 N \pi \rho_j}\right]    
\label{E_s}
\end{equation}   

This energy is the continuum  value (zero) plus a contribution
from the replacement of the line of charge distributions immediately
surrounding a given charge (which is included in the continuum energy) by the
self energy of the point charge which replaces it, and is not included in the
energetic calculation.  This is the charge in question $1/N$ times the
separation of the charges on the line,
$\log 1/N$.  One can also write this energy as an integral over the circle
of the form
\begin{equation}
 E_s = \frac{1}{2N} \int_0^{2\pi} ~d\theta~  \rho(\theta) 
\log\left[\frac{|F'(v)|}{ 2 N \pi\rho(\theta) }\right]   
\label{E_svar}
\end{equation}
Here $v$ is $e^{i\theta}$ and $\rho(\theta_j) $ is the same as $\rho_j$. 
This form of the energy is convenient for variational calculations. 

Note the factor of one half in front of equation (\ref{E_svar}).  It will be
important in what follows. 

The same discussion can be easily extended to the calculation of
the interaction energy.  Because of the logarithmic potential the result is
\begin{equation}
 E_{corr} = \frac{1}{2} \int_0^{2\pi} ~d\theta~  \rho(\theta) ~
 \int_0^{2\pi} ~d\mu~  \rho(\mu) ~
\Re \log[F(e^{i\theta} ) -F(e^{i\mu}  )]    
\label{E_corr}
\end{equation} 

\subsection{Variational Calculation}
To construct a variational principle, we take the two energies of equations
(\ref{E_svar}) and (\ref{E_corr})  add to them a
Lagrange multiplier term of the form
$$
 - \int_0^{2\pi} ~d\theta~  \rho(\theta)  \phi_0 
$$
designed to ensure that the total change is constrained to be unity. One
then sets the resulting variation with respect to $\rho$ to zero and finds
\begin{equation}
\phi_0 = \int_0^{2\pi} ~d\mu~  \rho(\mu) ~
\Re \log[F(e^{i\theta} ) -F(e^{i\mu}  )]  +
\frac{1}{2N}
\log \left[\frac{|F'(e^{i\theta})|}{2 N \pi e \rho(\theta)}\right]  
\label{phi02}
\end{equation}
The $e$ in the last term appears as a result of varying the density
inside the logarithm.  Note once more the factor $1/2$ in front of the
last term.   This equation is as far as we can tell, new.  

Equation (\ref{phi02}) looks like the statement that the electrical
potential on a conductor is constant.  However, that is not quite the
right interpretation.  To the requisite order ($1/N$) the right hand
side of equation (\ref{phi02}) is not the electrical potential.  The
electric potential is the somewhat similar expression in which the
symbols $e$ and $1/2$ just mentioned are replaced by unity, i.e.,
$$
\phi(\theta) = \int_0^{2\pi} ~d\mu~  \rho(\mu) ~
\Re \log[F(e^{i\theta} ) -F(e^{i\mu}  )]  +
\frac{1}{N}
\log \left[\frac{|F'(e^{i\theta})|}{2 N \pi \rho(\theta)}\right]  
$$
This actual potential, in contrast to $\phi_0$, varies in space. 

In equation (\ref{phi02}) the potential $\phi_0$ can be evaluated by
integrating the equation over all values of $\theta$.  The very same
arguments which we used in section \ref{complex} to perform the integration
over charges uniformly distributed on the circle can be used to simplify the
expression for 
$\phi_0$.  We find  
$$
2 \pi \phi_0 = \int_0^{2\pi} ~d \theta~   
\frac{1}{2N}
\log \left[\frac{|F'(e^{i\theta})|}{2 N \pi e \rho(\theta)}\right]  
$$   
We are looking for the lowest order terms which are of order $1/N$. Hence we
can set $\rho(\theta)$ to its lowest order value  $1/(2\pi)$. We can also
evaluate the integral over $\log F' $ by taking the contour to infinity.  In
the end we find
$$
2 \pi \phi_0 = \int_0^{2\pi} ~d \theta~   
\frac{1}{2N}
\log \left[\frac{1}{Ne~}\right]  
$$
so that, to order $1/N$ equation (\ref{phi02}) becomes
\begin{equation}
0 = \int_0^{2\pi} ~d\mu~  \rho(\mu) ~
\Re \log[F(e^{i\theta} ) -F(e^{i\mu}  )]  +
\frac{1}{2N}\log |F'(e^{i\theta})|  
\label{phi03}
\end{equation}   
which is our desired final result.

This equation is related to the integral equation for the charge
distribution in Ref.~\cite{Pom69a}. While our equation is a statement
of the constancy of an appropriate potential on the surface of the
conductor, the equation in Ref.~\cite{Pom69a} is a statement that the
tangential component of the electric field is zero at each charge
location, in an equilibrium configuration.

\section{Fourier space analysis} \label{sec:fourier}

We will analyze the integral equation (\ref{phi03}) using Fourier
analysis techniques. We define the Fourier coefficients of the
continuum density $\rho$ by
$$
\rho_k = \int_0^{2 \pi} \rho(\mu) e^{i \mu k} d \mu
$$ 
Following Pommerenke \cite{Pom67}, we also define
$$
s_k = \sum_{i = 1}^N w_i^k.
$$ 
where $w_i$ is the location of the $i$th charge in the minimum
energy configuration. The $s_k$ encode information about the locations
of the charges. For $k \ll N$, the $s_k$ reflect the smooth, large
scale behavior of the distribution of the charges. Thus, they should
correspond to the $\rho_k$ by
$$
s_k \approx N \rho_k = N \int_0^{2 \pi} \rho(\mu) e^{i k \mu} d \mu.
$$ For larger $k$, in particular for $k \sim N$, the discreteness of
the charges become important, and $s_k$ will in general be very
different from $\rho_k$.

We now rewrite the integral equation (\ref{phi03}) in terms of these
definitions. From (\ref{phi03}) we obtain
$$ 
\int_0^{2\pi} \rho(\mu) \Re \left[ \log[e^{i \theta} - e^{i \mu}] +
\log\left[\frac{F(e^{i\theta}) -F(e^{i\mu})}{e^{i \theta} - e^{i
\mu}}\right]\right] d \mu = - \Re \frac{1}{2N}\log [F'(e^{i\theta})]
$$ 
Substituting $\rho(\mu) = \frac{1}{2 \pi} \sum_k \rho_k e^{- i k
\mu}$, we obtain
$$
\frac{\rho_k}{k} + \sum_l a_{kl} \bar{\rho}_l = \frac{c_k}{2N},
$$ 
where, following Pommerenke \cite{Pom67}, we have defined the
matrix $a_{kl}$ and the vector $c_j$ by
$$ \log\left(\frac{F(w) - F(\xi)}{w - \xi}\right) = - \sum_{k =
1}^{\infty}\sum_{l = 1}^{\infty} a_{kl} w^{-k} \xi^{-l},
$$ 
and $c_j = -\sum_{k = 1}^{j-1} a_{k,j-k}$, so that
$$
\log F'(w) = -\sum_{k = 1}^{\infty} c_k w^{-k}.
$$ Solving this system of linear equations will yield the continuum
density $\rho$. Note that the above system is only meaningful for $k =
1,2,3,\ldots$. We know that the total charge is $1$, and this fixes
$\rho_0 = 1$.  We can also obtain information on the locations of the
charges form the correspondence between $\rho_k$ and $s_k$. For large
$N$, using the correspondence $s_k \approx N \rho_k$, we get
\begin{equation}
\frac{s_k}{k} + \sum_l a_{kl} \bar{s}_l = \frac{c_k}{2},
\label{eq:Pom}
\end{equation}
which is the same as the equation in Lemma 2 of Ref.~\cite{Pom67}. We
will henceforth refer to this as the Pommerenke equation. Note that
our analysis yields the (unexpected) conclusion that the equations
determining the $s_k$ have {\em no dependence} on $N$, the total
number of charges.

We can calculate $a_{kl}$ and $c_k$ for the ellipse and the
hypotrochoid using the explicit forms of $F$. In order to keep the
formulae simple, and avoid the appearance of complex roots of unity,
we will use the forms
$$
F(\xi) = \xi + \frac{c}{\xi^r},
$$ 
Therefore, the $c$ in this section is the same as the $c$ in the
previous section, if we set $\xi = e^{i \pi/2} w$ for the ellipse, and
$\xi = e^{i \pi} w$ for the hypotrochoid. We thus need to shift the
results with our choice of mapping function by $\pi/2$ and $\pi$
respectively to compare with the mapping functions in
Sec.~\ref{sec:real_space}.

A calculation for the ellipse yields
\begin{equation}
a_{kl} = \frac{c^k}{k} \delta_{kl}
\label{a:ellipse}
\end{equation}
and
\begin{equation}
c_k = \left\{ \begin{array}{cc} 0 & k \mbox{ odd} \\
\frac{2}{k}c^{k/2} & k \mbox{ even} \end{array} \right..
\label{c:ellipse}
\end{equation}

A similar calculation for the hypotrochoids shows that the energy
can be put in a similar form with
\begin{equation}
 a_{kl} = \left\{ \begin{array}{cc} 0 & k + l \neq 0 \bmod 3 \\ 0 &
k < p \mbox{ or } l < p, p = (k+l)/3 \\
\frac{c^p}{p}\frac{p!}{(k-p)!(l-p)!} & \mbox{ otherwise, } p = (k+l)/3 \end{array}
\right.
\label{a:hypotrochoid}
\end{equation}
and
\begin{equation}
c_k = \left\{ \begin{array}{cc} 0 & k \neq 0 \bmod 3 \\
\frac{3}{k}(2c)^{k/3} & \mbox{ otherwise. } \end{array} \right.
\label{c:hypotrochoid}
\end{equation}

\subsection{Solving the Pommerenke equation}

For the ellipse, the matrix $a_{kl}$ is diagonal. Consequently,
Eq.~(\ref{eq:Pom}) is easily solved to yield:
\begin{equation}
s_k = \left\{ \begin{array}{cc} 0 & k \mbox{ odd}, \\
\frac{c^{k/2}}{1 + c^k} = \frac{r_0^k}{1+r_0^{2k}} & k
\mbox{ even}. \end{array} \right.
\label{soln:ellipse}
\end{equation}
which is the solution presented in \cite{Pom67}. Note that for this
solution, $s_k$ is real for all $k$. This is equivalent to the
statement that there is no {\em symmetry breaking}. Thus this approach
cannot capture the symmetry breaking for the ellipse, that we see in
the numerics. 

\begin{figure}
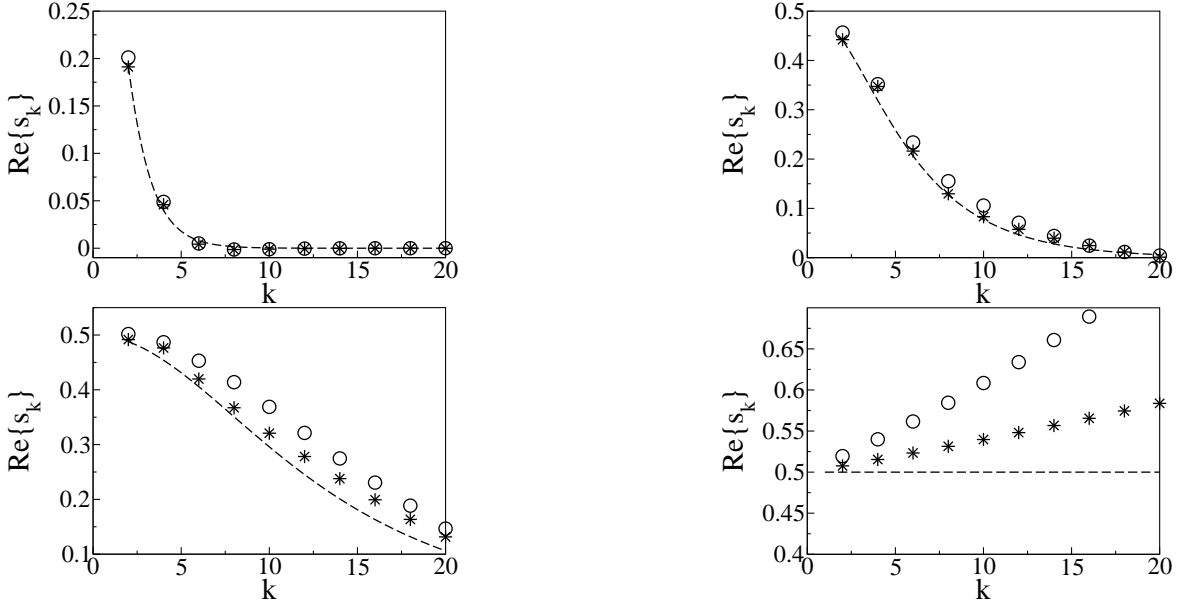

\centering
\includegraphics[width=6cm]{sk-e-100-0.2.eps}
\hfill
\includegraphics[width=6cm]{sk-e-100-0.6.eps}\\
\includegraphics[width=6cm]{sk-e-100-0.8.eps}
\hfill
\includegraphics[width=6cm]{sk-e-100-1.eps}
\caption{$s_k$ as a function of $k$ for the ellipse, for c=0.2, 0.6, 0.8 and 1 
respectively; the circles are the numerics obtained with $N=40$, 
the stars with $N=100$,  while the dashed line is expression 
\ref{soln:ellipse}. As expected the agreement is good for small values of 
$c$, when the symmetry is preserved. The deviations are therefore stronger for 
$N=40$ than for $N=100$.}
\label{sk-ell}
\end{figure}

This result is compared to numerical calculations in Fig.\
\ref{sk-ell}, which is also done with the map $F(w) = w + c/w$.  The
circles are the results obtained for $N=40$, while the stars
correspond to $N=100$. As expected, the agreement is good for small
enough values of $c$, for which the symmetry is unbroken. Moreover the
agreement is better for $N=100$ than for $N=40$. This is in agreement
with the expectation that deviation of the solution
(\ref{soln:ellipse}) from the true $s_k$ is $O(k^2/N)$ \cite{Pom67}.

We will now consider the hypotrochoid. We first enumerate the
consequences of the three-fold symmetry as well as the reflection symmetry of the hypotrochoid. We get,
\begin{enumerate}
\item $s_k = 0$ for $k \neq 0 \bmod 3$ is a reflection of
the symmetry of the underlying hypotrochoid under rotations by $2
\pi/3$. This symmetry is preserved in our numerically obtained
solutions if the number of charges $N$ is a multiple of 3, and is
broken otherwise. 
\item $s_k$ is real as a consequence of the symmetry of the
hypotrochoid under reflections about the real axis. This symmetry is
broken in our numerical solutions for any $N$, provided that $c$ is
sufficiently close to 1/2.
\end{enumerate}
If there is a unique minimizing configuration for the electrostatic
energy of the hypotrochoid, the symmetries of the problem imply that
$s_k = 0$ if $k \neq 0 \bmod 3$ and $s_k$ is real for all $k$. These
features are also reflected in the (approximate) $s_k$ obtained by
solving the Pommerenke equation. Since the Pommerenke equation is
independent of $N$, we cannot investigate the $N$ dependent features,
like symmetry breaking for the hypotrochoid, using this approach.

We will now solve the Pommerenke equation for the hypotrochoid. If $c
< 1/2$, $c_k$ and the $a_{kl}$ decay exponentially as $k$ and $l$ get
large. Therefore, it is reasonable to try the series solution
\begin{equation}
s_k = \frac{k c_k}{2} - \sum_l (k a_{kl}) \frac{l \bar{c}_l}{2} 
+ \sum_j\sum_l (k a_{kl}) (l a_{lj}) \frac{j \bar{c}_j}{2} + \ldots.
\nonumber
\end{equation}
This series converges (not just for the hypotrochoid, but for any
admissible $a_{kl}$), as proved by Pommerenke \cite{Pom67}.

Let $s_k^{(i)}$ denote the $i$-th term in the
series, so that
\begin{eqnarray}
s^{(1)}_k & = & \frac{k c_k}{2} \nonumber \\
s^{(i+1)}_k & = & - \sum_l k a_{kl} \bar{s}^{(i)}_l, \quad 
\quad i = 2,3,\ldots\nonumber
\end{eqnarray}
Eqs.~(\ref{a:hypotrochoid}) and (\ref{c:hypotrochoid}) together give
\begin{align}
s^{(1)}_k & =  0, \quad k \neq 0 \bmod 3 \nonumber \\
s^{(1)}_{3p} & =  \frac{3}{2}(2 c)^{p} = \frac{3}{2}r_0^{3p}  \nonumber \\ 
s^{(i+1)}_k & =  -
\sum_{p = \left[\frac{k+1}{2}\right]}^{k} \left(\frac{k
c^p}{p}\right) \frac{p!}{(k-p)!(2p - k)!}  \bar{s}^{(i)}_{3p-k}, \quad i = 1,2,3,\ldots
\label{eq:series_soln}
\end{align}
Since $3p -k \neq 0 \bmod 3$ unless $k = 0 \bmod 3$, it follows that
$s^{(i)}_k = 0$ if $k \neq 0 \bmod 3$ for all $i$. Setting $k = 3m, p
= m+n$ and replacing the sum over $p$ by a sum over $n = p -m$, we get
\begin{eqnarray}
s^{(i+1)}_{3m} & = &  -
\sum_{n = \left[\frac{m+1}{2}\right]}^{2 m} 
\left(\frac{3 m}{m+n}\right) \frac{(m+n)!}{(2m-n)!(2n - m)!} 
\left(\frac{r_0^3}{2}\right)^{m+n}\bar{s}^{(i)}_{3n} 
\quad \quad i = 1,2,3,\ldots \nonumber \\
& = & - 3 \left(\frac{r_0^3}{2}\right)^m 
\sum_{n = \left[\frac{m+1}{2}\right]}^{2 m}
\left(\frac{m}{n+m}\right) 
\left(\begin{array}{c} m+n \\ 2m-n \end{array}\right) 
\left(\frac{r_0^3}{2}\right)^n \bar{s}^{(i)}_{3n}. \nonumber
\end{eqnarray}

We can evaluate these sums using the saddle point method. Using
Lemma~\ref{lem:saddle} in Appendix~\ref{sec:apndxa}, and $s^{(1)}_{3m} =
\frac{3}{2}r_0^{3m}$, we obtain the asymptotic expressions
$$
s^{(i)}_{3m} \sim (-1)^i \frac{3 \beta_i^m}{2}, \quad i = 1,2,3,\ldots
$$
where 
\begin{eqnarray}
\beta_1 & = & r_0^3 \nonumber \\
\beta_{i+1} & = & \frac{\left[(r_0^3 \beta_i + \sqrt{8 r_0^3 \beta_i + 
r_0^6 \beta_i^2})\right]^3}{64 \beta_i}, \quad i = 1,2,3,\ldots \nonumber
\end{eqnarray}
and all corrections to this asymptotic result are $O(1/m)$.

At the critical parameter value $c = \frac{1}{2}$, $r_0 = 1$. In this
case, $\beta_1 = 1$ and
$$
\beta_{i+1} = \frac{\left[(\beta_i + \sqrt{8 \beta_i + 
\beta_i^2})\right]^3}{64 \beta_i}
$$ 
implies that $\beta_i = 1$ for all $i$. Thus, the series solution
breaks down in the critical case, as the solutions $s^{(i)}_k$ do not
decay with $i$.

\begin{figure}
\centering
\includegraphics[width=6cm]{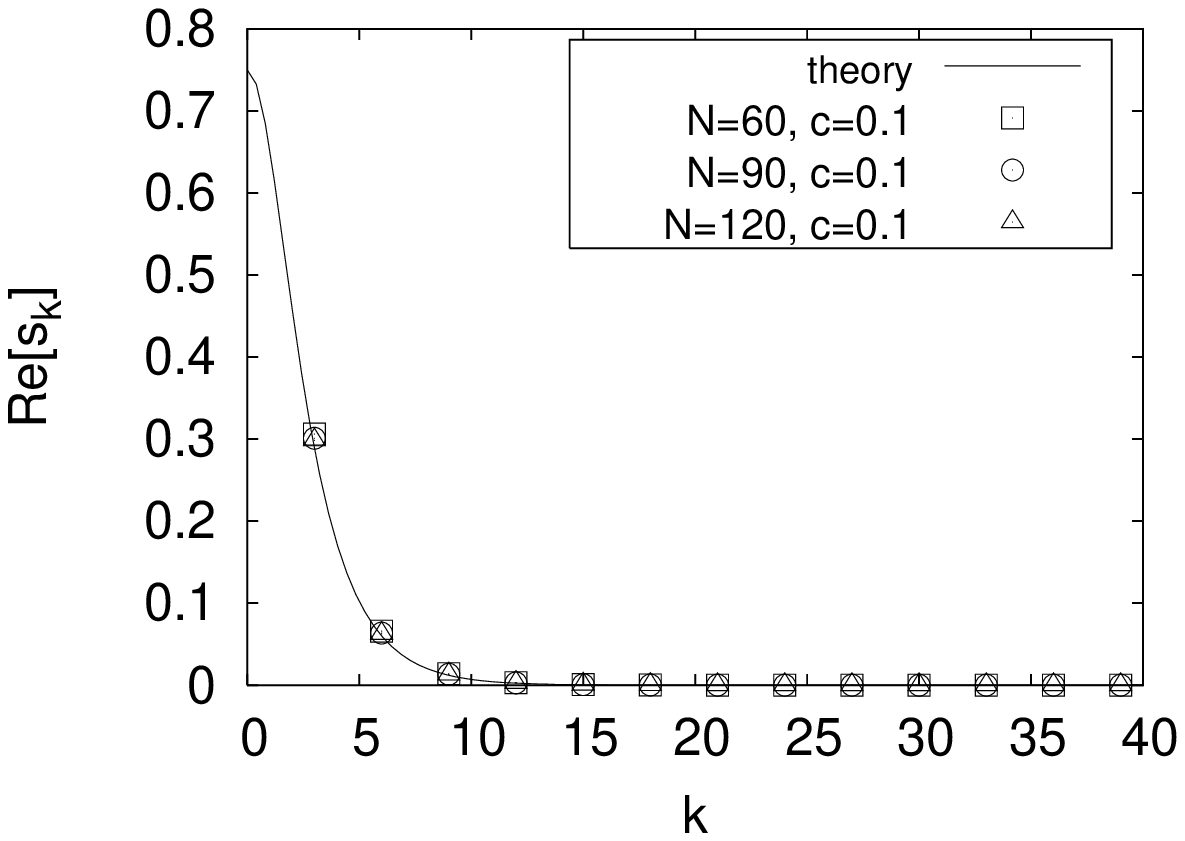}
\hfill
\includegraphics[width=6cm]{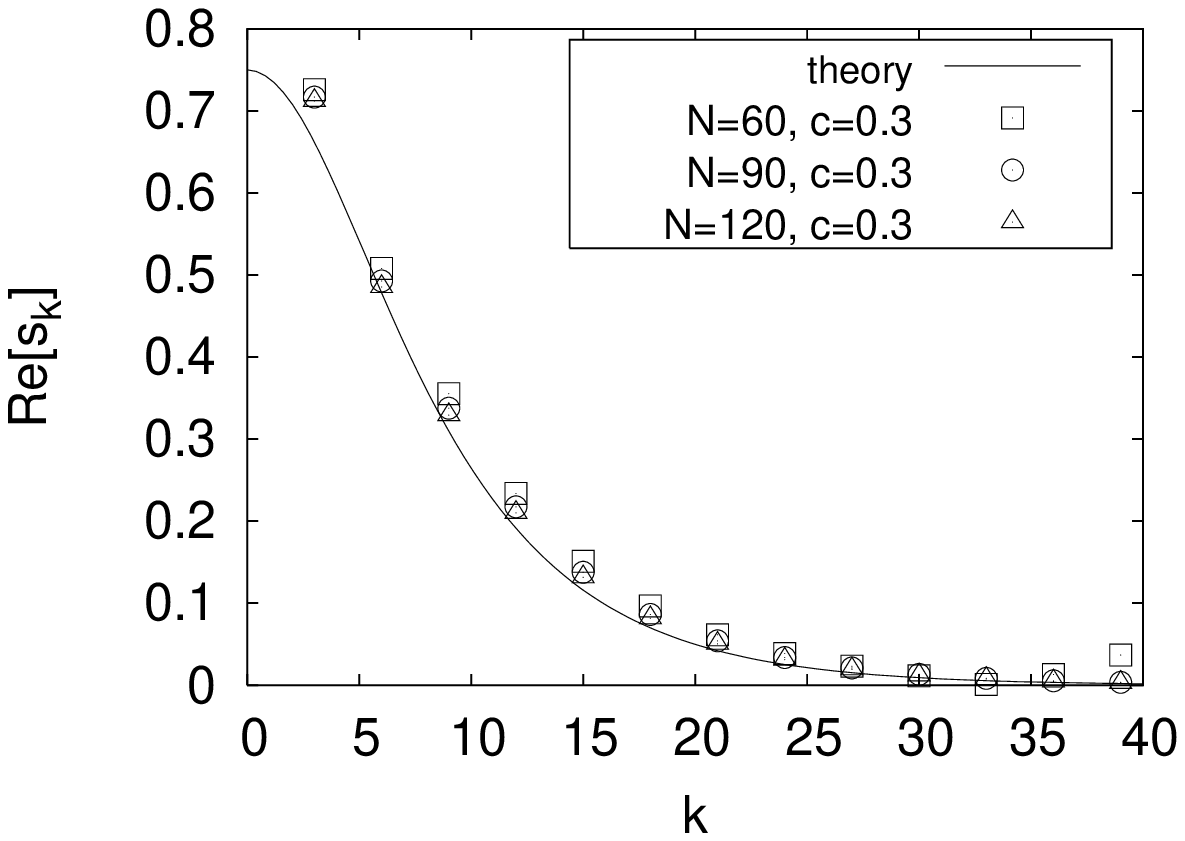}\\
\includegraphics[width=6cm]{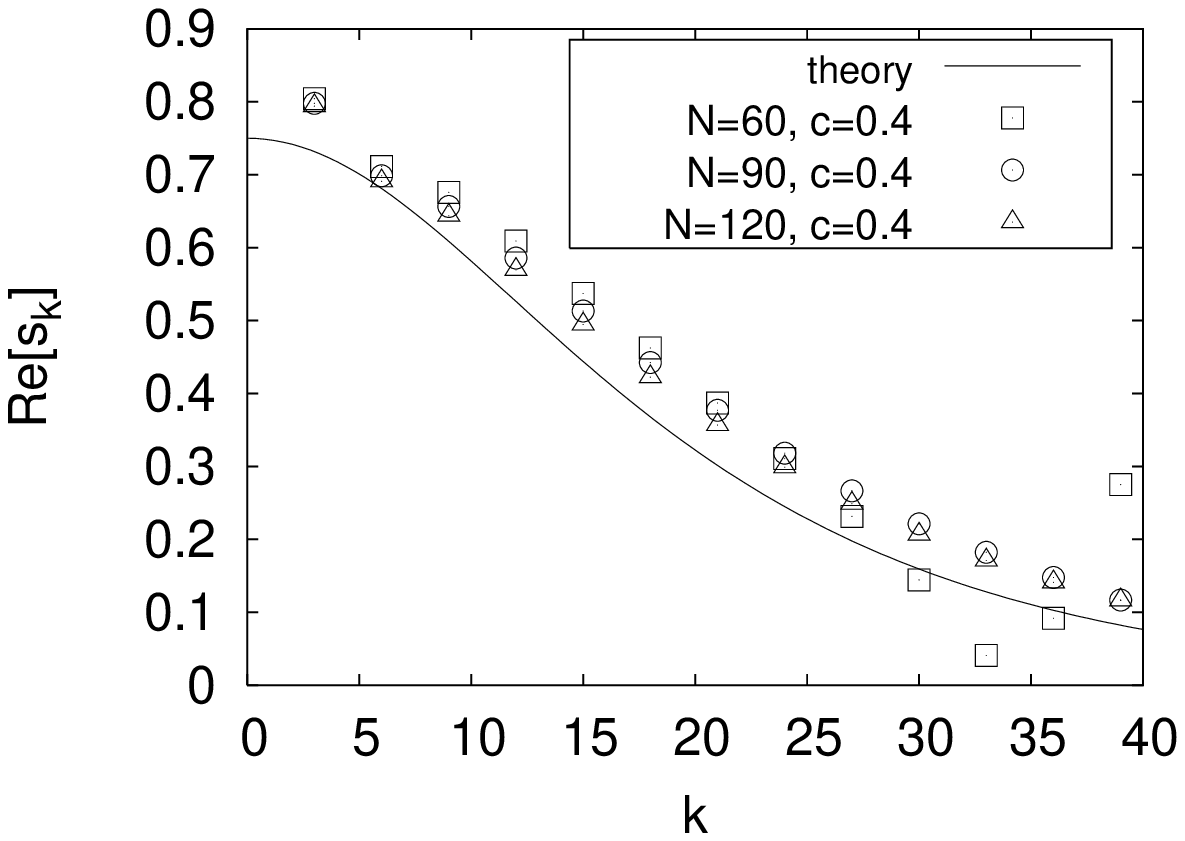}
\hfill
\includegraphics[width=6cm]{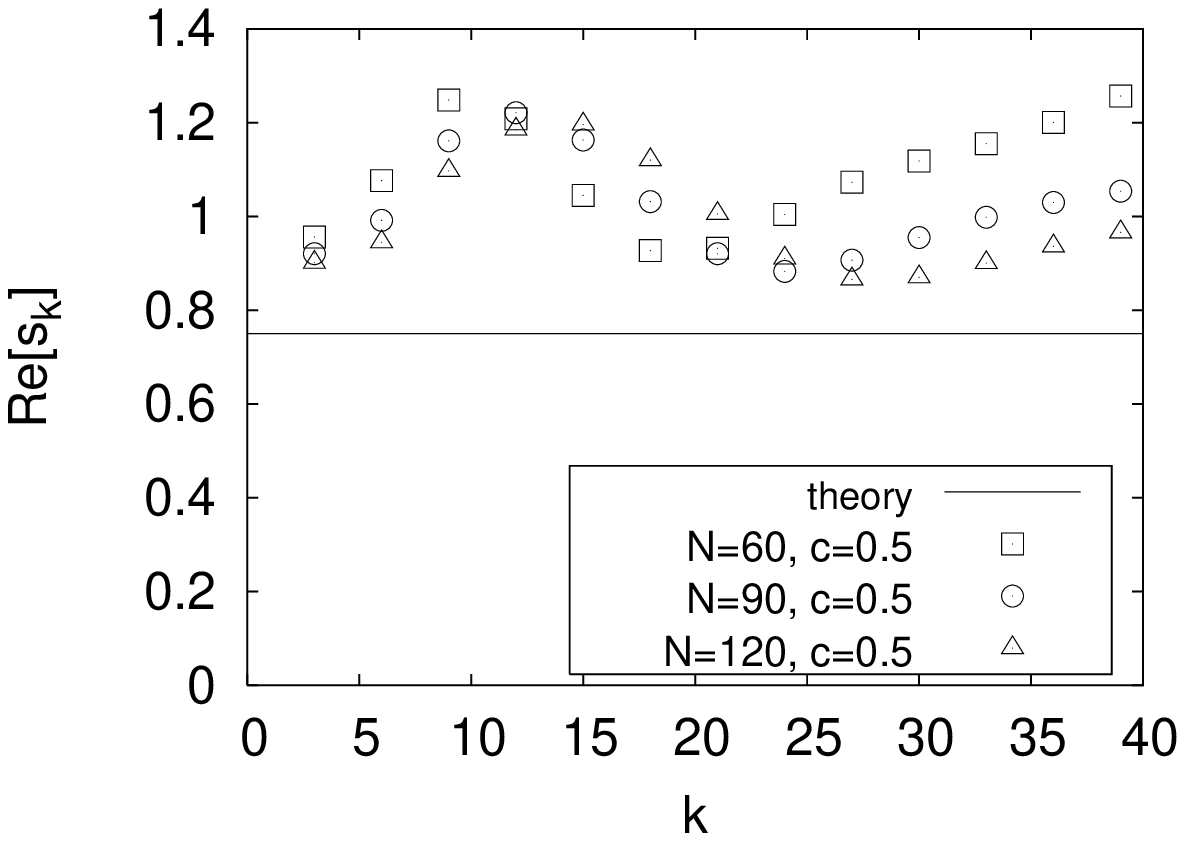}
\caption{$s_k$ as a function of $k$ for the hypotrochoid, 
for c=0.1, 0.3, 0.4 and 0.5 
respectively. As expected the agreement is good for small values of 
$c$, when the symmetry is preserved.}
\label{sk-hypo}
\end{figure}

For parameter values close to the critical value, $c =
\frac{1}{2}e^{-\epsilon}$, we have
$$
r_0^3 = 2c = e^{-\epsilon} \approx 1 - \epsilon
$$ If $\epsilon$ is sufficiently small, then we have $\beta_i$ is
close to one for the first few values of $i$. We can study the
dependence of $\beta_i$ on the parameter $\epsilon$ by setting
$\beta_i = e^{-t_i \epsilon + O(\epsilon^2)} = 1 - t_i \epsilon +
O(\epsilon^2)$. Linearizing the relation between $\beta_{i+1}$ and
$\beta_i$ close to the fixed point $\beta = 1, r_0 = 1$, we get
$$ 
\beta_{i+1} = 1 - (t_i+2) \epsilon + O(\epsilon^2) + \ldots
$$
This gives
$$
t_{i+1} = t_i + 2.
$$
Combining this with $t_1 = 1$, we get
$$
t_i = (2i-1) 
$$ It is clear that this analysis is valid long as $t_i \epsilon \ll
1$, {\em i.e.}, for $i \lesssim \epsilon^{-1}$. Beyond this value, we
can show that the $\beta_i$ decay rapidly. Using this expression for
the $\beta_i$, we can re-sum the series solution for $s_k$ to obtain
$$
s_{3m} = \frac{3}{2} \frac{r_0^{3m}}{1 + r_0^{6m}} + O\left(\frac{1}{m}\right).
$$
The complete solution is
\begin{equation}
s_k = \left\{ \begin{array}{cc} 0 & k \neq 0 \bmod 3, \\
\frac{3}{2}\frac{(2c)^{k/3}}{1 + (2c)^{2k/3}}(1 + O(k^{-1})) \approx
\frac{3}{2}\frac{r_0^k}{1+r_0^{2k}} & k = 0 \bmod 3. \end{array}
\right.
\label{soln:hypotrochoid}
\end{equation}
Note the similarity with the solution in Eq.~(\ref{soln:ellipse}).
The agreement of result (\ref{soln:hypotrochoid}) with numerics is
tested in Fig.\ \ref{sk-hypo}. We use the map $F(w) = w - c/w^2$, but
the quantities plotted are $(-1)^k s_k$, which is the same thing as
using the map $F(w) = w + c/w^2$. For small $c$ and for $k \ll N$, the
agreement is very good. However, the agreement is rather poor if $c
\sim 1/2$, indicating that the integral equation is no longer valid as
we approach the singular limit $c \rightarrow 1/2$.

\subsection{The minimum energy configuration of the charges}

Our goal is to find the locations of the $N$ charges on the ellipse
and hypotrochoid that minimize the electrostatic energy $E_N$. The
results from the previous section tell us that the locations $w_i$ on
the unit circle that give the appropriate locations on the ellipse and
hypotrochoid by $z_i = F(w_i)$ are such that
\begin{equation}
\sum_{i=1}^n w_i^k \approx \left\{ \begin{array}{ll} \frac{r_0^k}{1 + r_0^{2k}} &
k \mbox{ even, on ellipse}. \\
 \frac{3}{2}\frac{r_0^k}{1+r_0^{2k}} + O\left(\frac{1}{k}\right)& k = 0 \bmod 3, \mbox{ on hypotrochoid}.\\
0 & \mbox{ otherwise} \end{array}
\right. 
\label{eq:minimizers}
\end{equation}
for $1 \leq k \ll N$. The problem of finding the locations $w_i$
reduces to the problem of solving the above set of equations.

As Pommerenke shows in Ref.~\cite{Pom67}, the locations of the charges
are given in terms of the solution $s_k$ to the Pommerenke equation by
\begin{equation}
\theta_{j} \approx \theta_0 + \frac{2 \pi j}{N} + \frac{1}{N} \left[ \psi\left(\theta_0 + \frac{2 \pi j}{N}\right) - \psi(\theta_0)\right] 
\label{eq:phifuction}
\end{equation}
where
$$ 
\psi(\theta) =  i\sum_{k = 1}^{\infty} \left[\frac{s_k}{k}
e^{-i k \theta} - \frac{\bar{s}_k}{k} e^{i k \theta}\right]
$$ 
and $\theta_0$ is an initial angle that is not determined by this
analysis. 

For the ellipse and the hypotrochoid, we get the series
$$ \psi(\theta) = -\sum_{k = 1}^{\infty}
\frac{1}{k} \frac{q^k}{1 + q^{2k}} \sin(m k \theta),
$$ 
where $m = 2,3$ for the ellipse and hypotrochoid respectively, and
$q = r_0^m$. This series can be summed  using the formula
$$ 
\mathrm{am}(u) = \frac{\pi u}{2K} + 2 \sum_{n = 1}^{\infty}
\frac{1}{n} \frac{q^n}{1+q^{2n}} \sin \frac{n \pi u}{K}.
$$
(See p. 277, Ryshik and Gradstein
\cite{tables_spi}) to yield
$$
\psi(\theta) =  \frac{m \theta}{4} - 
\frac{1}{2}\mathrm{am}\left(\frac{m K \theta}{\pi}\right).
$$
$K = K(k)$ is the complete Jacobi elliptic function of the
first kind \cite{whittaker_watson}, and the modulus $k$ is determined
by requiring that $q$ be the nome \cite{Abram_Stegun},
$$
q = \exp\left[-\frac{\pi K'(k)}{K(k)}\right],
$$ 
where $K'(k)$ is the complementary complete elliptic function of
the first kind, $K'(k) = K(\sqrt{1-k^2})$.

We are particularly interested in situations close to the critical
value of $c$. In this case, $q = e^{- \epsilon} \approx 1 - \epsilon$
is close to 1. From this, we infer that $K'/K \approx 0$, {\em i.e}
$k$ is close to 1. We will now work out the asymptotics close to $k =
1, q = 1$.

If $k = 1 - \delta$, $k^2 \approx 1 - 2 \delta$ and $k' = \sqrt{1 -
k^2} \sim \sqrt{2 \delta}$, and we have
\begin{eqnarray}
K(k) & = & \int_0^{\pi/2} \frac{d x}{\sqrt{1 - k^2 \sin^2x}} \nonumber
\\ & = & \int_0^{\pi/2} \frac{dx}{\sqrt{2 \delta + \cos^2(x)}} +
\xi(\delta) \nonumber \\ 
& = & - \frac{1}{\sqrt{2}} \log(\delta) + \xi'(\delta) \nonumber
\end{eqnarray}
where we have extracted the dominant singularity of the integral in
the second line, and $\xi,\xi'$ are functions that remain bounded as
$\delta \rightarrow 0$. We also have 
$$ 
K' \sim \int_0^{\pi/2}\frac{dx}{\sqrt{1 - 2 \delta \sin^2x}} =
\frac{\pi}{2} + O(\delta).
$$ 
Combining this with the previous result, and the definition of the
nome $q$, we see that
$$
q \approx \exp\left[\frac{\pi^2}{\sqrt{2} \log(\delta)}\right].
$$
Rearranging this gives the result
$$ 
\frac{\pi^2}{\sqrt{2}} - \log(q) \log(\delta) \rightarrow 0 \quad
\text{as } \delta \rightarrow 0.
$$ 
Using $\log(q) = - \epsilon$, we see that $\delta \sim \exp[-
\pi^2/(\sqrt{2} \epsilon)]$, so that $K \sim - \log{\delta}/\sqrt{2} 
\sim \pi^2/(2 \epsilon)$. 

\begin{figure}
\centering
\includegraphics[width=\hsize]{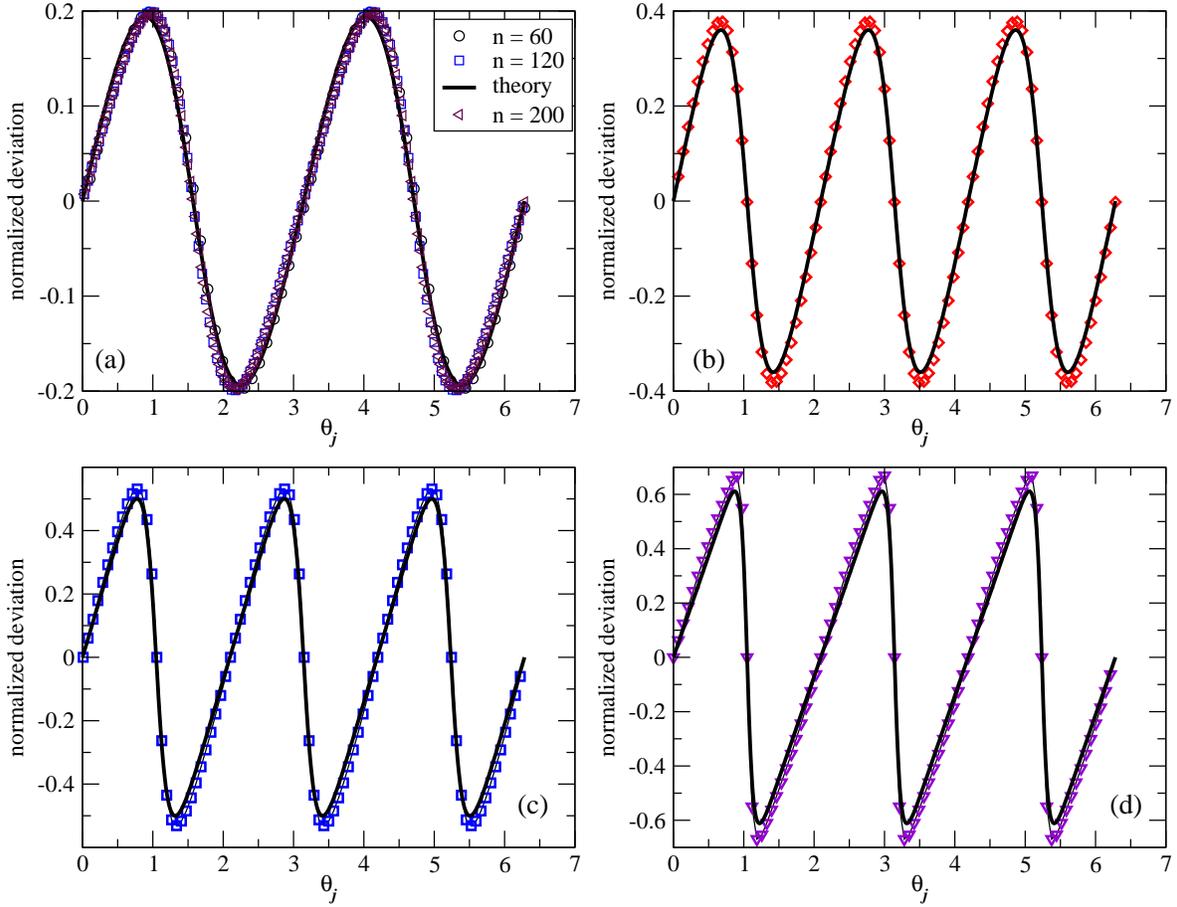}
\caption{A comparison between the theoretical prediction and the
actual locations of the charges in the minimum energy
configuration. The normalized deviation from the nominal position
$\theta_j = 2 \pi j/N$ is plotted. (a) shows the deviations for an
ellipse with $c = 0.2$. (b),(c) and (d) depict the charge placement on
a hypotrochoid with $c = 0.2, 0.3$ and $0.4$ respectively. The solid
lines are the theoretical prediction.}
\label{fig:phi_function}
\end{figure}

Fig.~\ref{fig:phi_function} is a comparison between the results of
this calculation, and the numerically observed displacements of the
charges form the positions that would correspond to a uniform
distribution on the circle. The normalized deviation function
\begin{equation}
\psi(\theta) =  \frac{m \theta}{4} - 
\frac{1}{2}\mathrm{am}\left(\frac{m K \theta}{\pi}\right).
\label{eq:placement}
\end{equation}
corresponds to the following two changes relative to the uniform
distribution on the unit circle --
\begin{enumerate}
\item The ``charge'' density in the smooth regions is increased by an
amount $\delta \rho/\rho = m/(4N)$. 
\item The charge moves away from the vicinity of the cusps, and this
depletion region has a width $\pi/(mK) \sim 2 \epsilon/(m \pi)$ around
the the angles corresponding to the cusps.
\end{enumerate}
Note that Figs.~\ref{fig:phi_function} (b), (c) and (d) are obtained
with the mapping $F(w) = w - c/w^2$ and the solid lines are therefore
given by the ``shifted'' function $\psi(\theta + \pi)$.
\section{The singular limit} \label{sec:singular}

In this section, we investigate various aspects of the Coulomb problem
on domains that are either singular, {\em i.e.}, have cusps or
corners, or are close to being singular. We begin by first describing
some of the numerical observations, which motivate the theoretical
analysis that follows.

\subsection{Ellipse, hypotrocho\"id and related curves.}

\subsubsection{Symmetry breaking}
As previously mentioned, an interesting feature of the charge
placement on the ellipse and on the hypotrocho\"id 
is that the minimum energy configurations
break the symmetry of the problem in certain ranges of $c$ and $N$.
For a given number of charges $N$, if the value of the parameter $c$ is
larger than a given threshold $c^*$ which depends on $N$, the mirror
symmetry is broken around each pointy region.  The larger $N$, the
closer the threshold is to the critical value, {\em i.e}, $c^*(N)
\rightarrow 1$ for the ellipse and $c^*(N) \rightarrow \frac{1}{2}$
for the hypotrochoid as $N \rightarrow \infty$.
%

\begin{figure}[htbp]
\centering
\includegraphics[width=8cm]{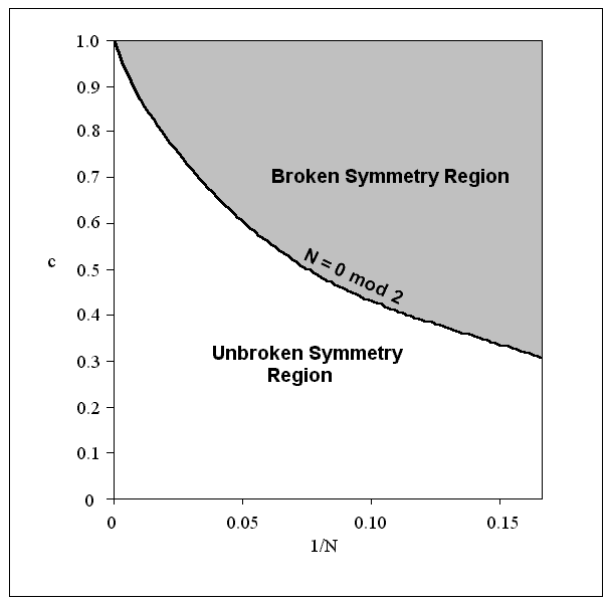}
\caption{Phase diagram for the ellipse}
\label{PDell}
\end{figure}

\begin{figure}[htbp]
\centering
\includegraphics[width=8cm]{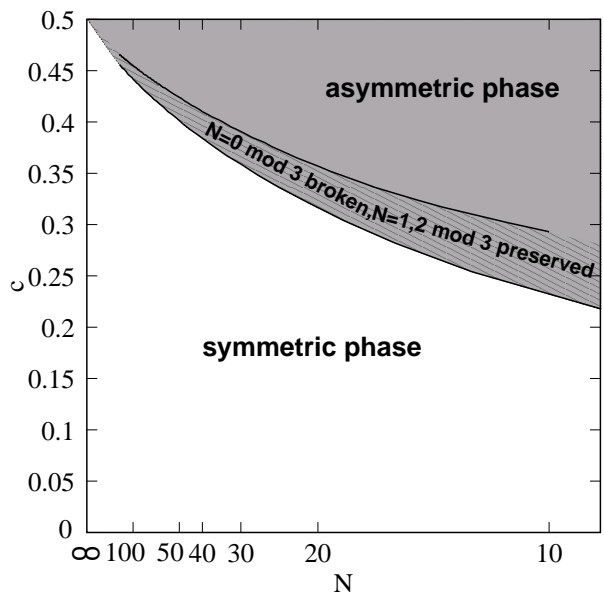}
\caption{Phase diagram for the hypotrochoid}
\label{PDhypo}
\end{figure}

This transition can be described by a phase diagram where $c^*$ 
is plotted versus $N$: the 
case of the ellipse is presented in Fig.\ \ref{PDell} and the case of the 
hypotrochoid in Fig.\ \ref{PDhypo}.

\begin{figure}[htbp]
\includegraphics[width=9.5cm]{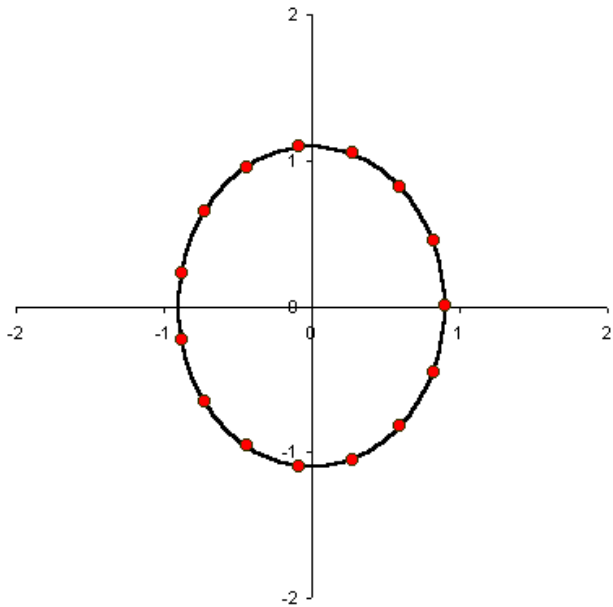}
\hfill
\includegraphics[width=9.5cm]{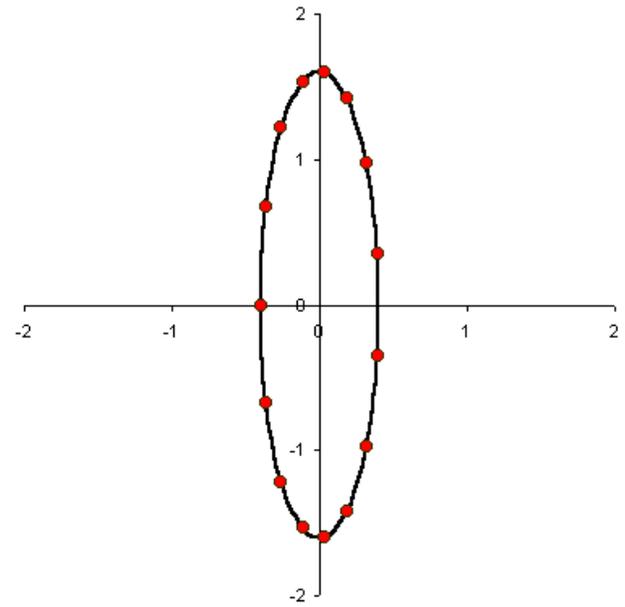}
\caption{ Placement of 15 charges on two different  ellipses, 
$c=0.1$ and $c=0.7$ respectively.  Notice the
breaking of the parity symmetry in part b.} 
\label{unbroken_odd}
\end{figure}

On both phase diagrams, two different cases should be distinguished, depending 
on the number of charges $N$. Let us first consider the case of the ellipse. 
If 
$N$ is even, in the unbroken symmetry phase, the charge configuration will be 
symmetric versus both the vertical and the horizontal axis. Both symmetries 
break when $c>c^*$ (see Figure \ref{ellipse1}). 
On the contrary, for odd $N$, there is no symmetry across the long axis, while there is always symmetry across the short axis (see Figure \ref{unbroken_odd}).  In this case, there is no transition of any sort.
The value $c^*$ at which the symmetry breaks for even $N$ forms a curve (see Figure \ref{PDell}).

\begin{figure}[htbp]
\includegraphics[width=9.5cm]{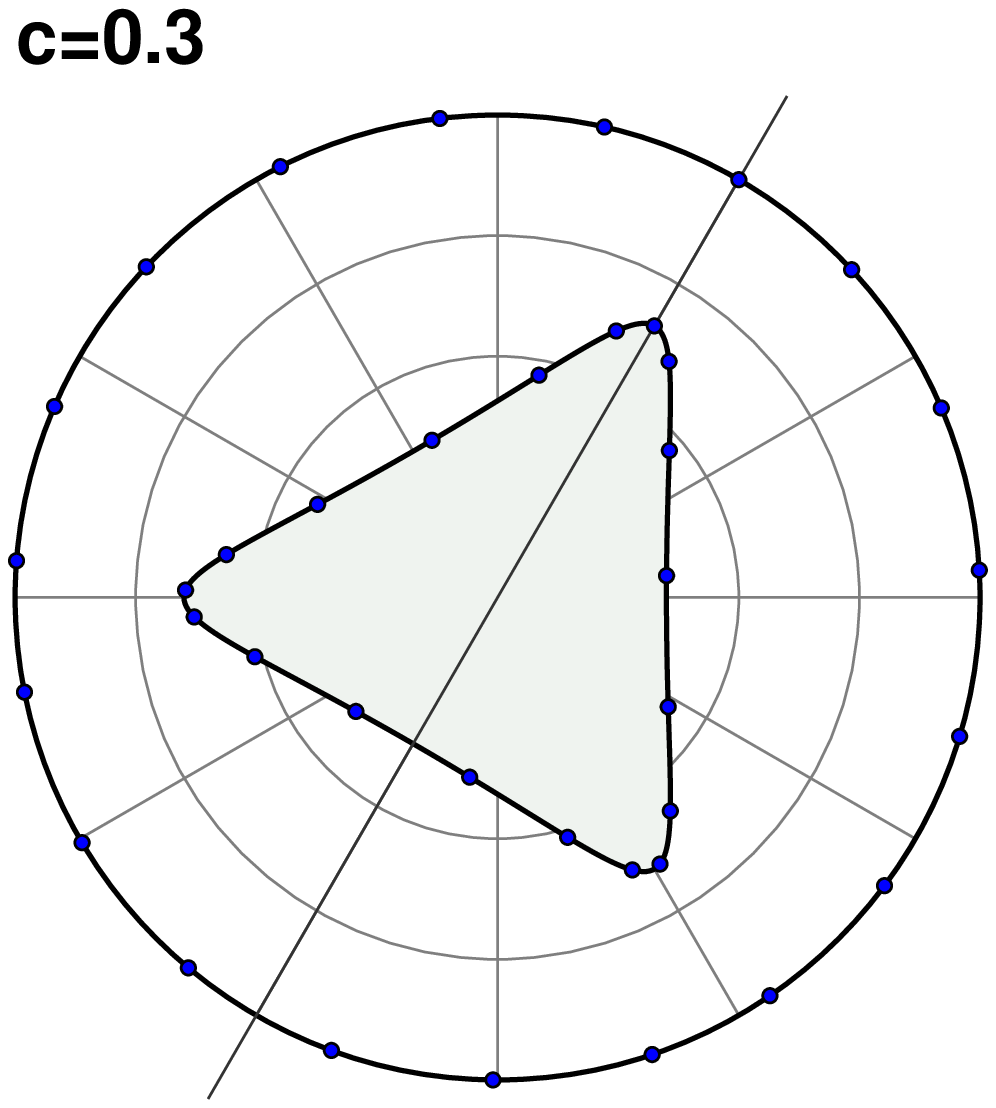}
\hfill
\includegraphics[width=9.5cm]{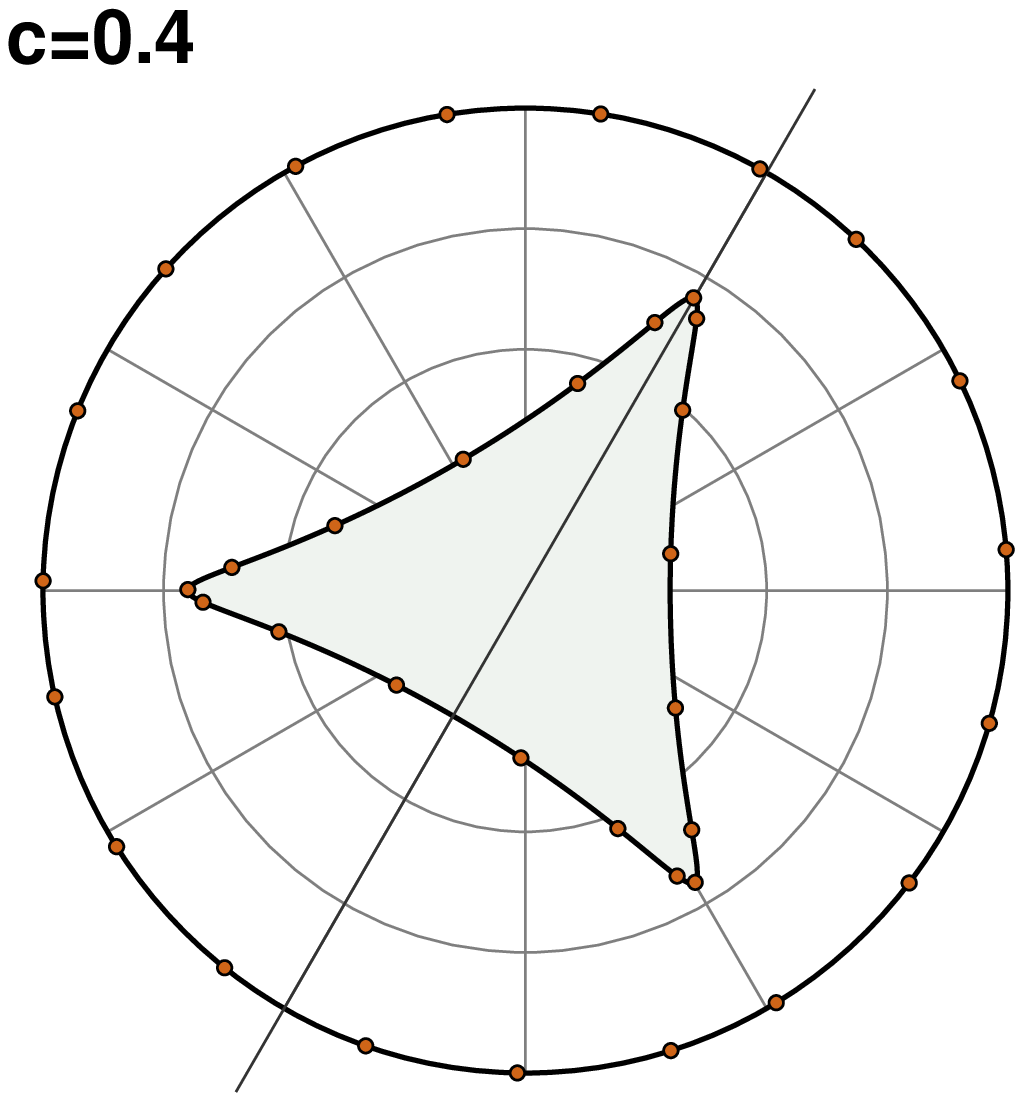}
\caption{ Placement of 19 charges on two different hypotrochoids, 
$c=0.3$ and $c=0.4$ respectively.  Notice the
breaking of the parity symmetry in part b.} 
\label{hypo2}
\end{figure}

Similarly, in the case of the hypotrochoid, the case $N=0$ mod 3 has to be 
distinguished from the case $N \ne 0$ mod 3. In the first case, a symmetric 
charge configuration respects the three mirror symmetries, one at each pointy 
region (see Figure \ref{hypo1}). In the case $N \ne 0$ mod 3, only 2 out of 
the 3 branches of the hypotrochoid are identical, with one charge more or less 
than the third branch. 
Therefore, only one mirror symmetry is respected in this 
case (see Figure \ref{hypo2}). 
The boundary between the broken and the unbroken symmetry phases is 
different for $N=0$ mod 3 and for $N \ne 0$ mod 3 (see Figure \ref{PDhypo}).

\subsubsection{Energy dependence on $N$}

\begin{itemize}
 
\item{Line segment:}
As pointed out previously, a useful quantity to characterize the placement 
of charges on a given  curve is the energy as a function of 
the number of charges. 
The case of the line 
segment $\Gamma_2=[-2,2]$ has been treated analytically in literature 
\cite{Pom64}. 
The energy dependence on $N$ is given by
$$E_{\Gamma}^N=- \frac{\ln N}{2 N}-\frac{\ln 2}{2 N}-\frac{\ln N}{8 N^2}-\frac{A}{2 N^2},$$
where $A$ is a constant. The first term is the self-energy. 
The first correction, or correlation energy, 
goes as $N^{-s}$ with $s=1$ in this case.

Other curves presenting this kind of $N$-dependence for the correlation 
energy are the star shape curves defined 
by the conformal mapping $F(w)=z(1+c w^{-m})^{2/m}$\cite{Pom64}, with $c=1$, 
where $w=e^{i\theta}$ is on the unit circle and 
$m$ is the number of star branches. 
Figure \ref{star3-1} shows the case $m=3$. 
The dependence of the correlation energy on $N$ is shown 
in Fig.\ \ref{star3-E1-1}.  
Notice that the charges are grouped near the points and that the density is lower at the center.

\begin{figure}[htbp]
\centering
\includegraphics[width=5cm]{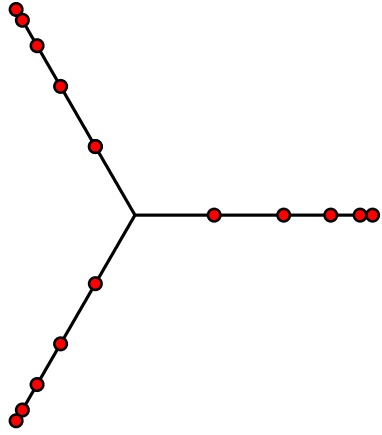}
\caption{Placement of 15 charges on a star shape curve with 3 branches for $c=1$.}
\label{star3-1}
\end{figure}

\begin{figure}[htbp]
\centering
\includegraphics[width=8cm]{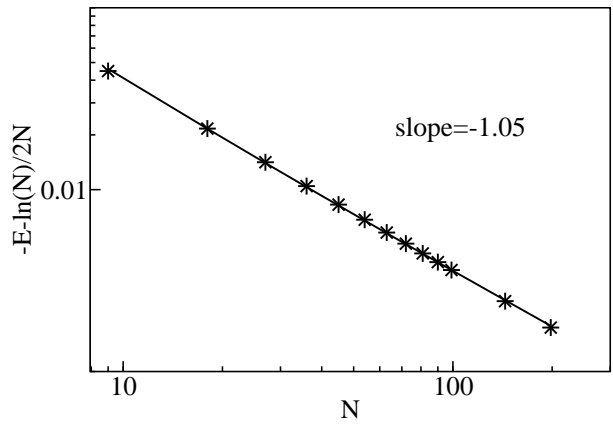}
\caption{Dependence of the correlation energy on $N$ for the 
star shape curve with 3 branches. 
The slope shown for the data is consistent with the value -1.0.
 }
\label{star3-E1-1}
\end{figure}

\item{Ellipse:}
The line segment $\Gamma_2$ is the singular limit, $c=1$, 
of the ellipse defined by the 
conformal map $F(w)=w-\frac{c}{w}$. For a smooth curve, $c<1$, 
the correlation energy goes as 
$N^{-s}$ with $s=2$ (see Fig.\ \ref{crossell}). 
See  the work of Pommeranke \cite{Pom67} where it is shown that s=2 for any smooth curve.  
In the case $c$ close to but smaller than 1, we observe a crossover 
as a function of the number of charges $N$: 
for smaller values of $N$, $s=1$, while 
for larger values of $N$, $s=2$, as shown in Fig.\ \ref{crossell}.

\begin{figure}[htbp]
\centering
\includegraphics[width=8cm]{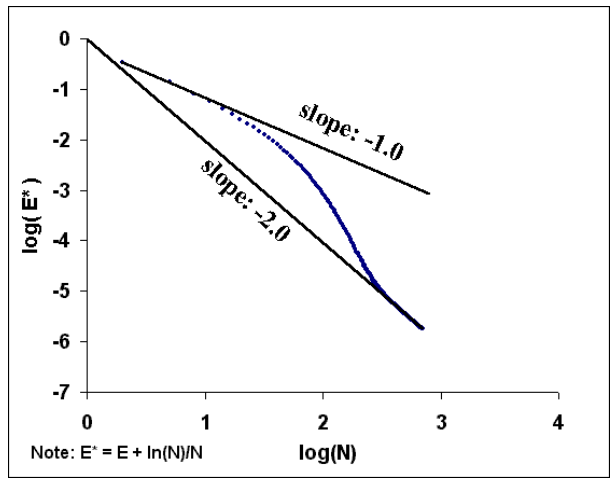}
\caption{Dependence of the correlation energy on $N$ for the 
for the ellipse, with $c=0.975$ }
\label{crossell}
\end{figure}

The same $N$ dependences are observed for the star shape curves for $0<c<1$ 
(see Fig.\ \ref{star3-0.5}). 
The different situations are depicted in Fig.\ \ref{star3-E1-0.5}.

\begin{figure}[htbp]
\centering
\includegraphics[width=5cm]{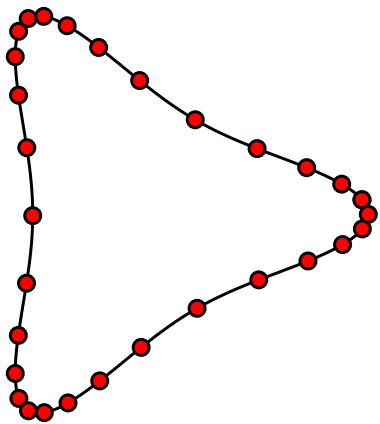}
\caption{Placement of 30 charges on a star shape curve with 3 branches, 
with $c=0.5$.}
\label{star3-0.5}
\end{figure}

\begin{figure}[htbp]
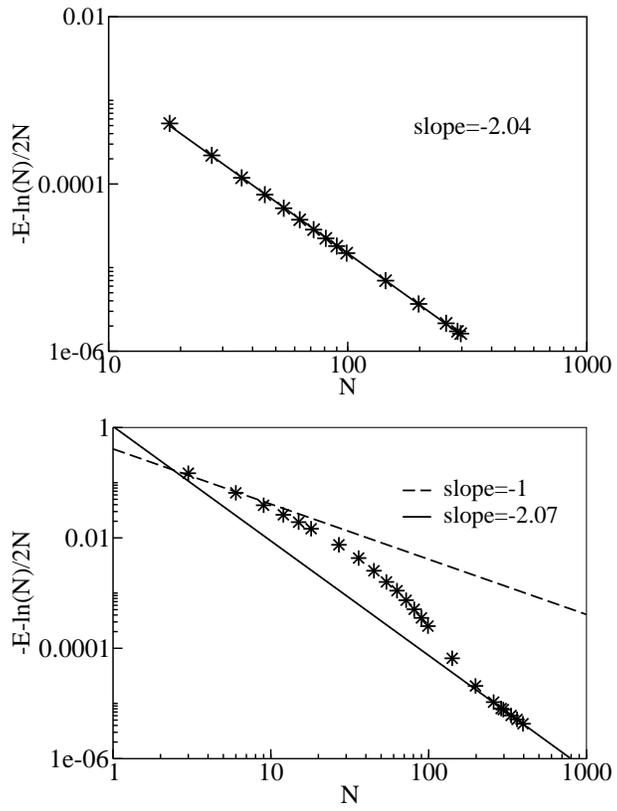

\centering
\includegraphics[width=8cm]{star3-E1-0.5.eps}
\\[0.25cm]
\includegraphics[width=8cm]{star3-E1-0.95.eps}
\caption{Dependence of the correlation energy on $N$ for the 
star shape curve with 3 branches, for $c=0.5$ and $c=0.95$.
The slopes of the line is consistent with the theoretical values of -1.0 and -2.0.
}
\label{star3-E1-0.5}
\end{figure}

\item{Hypotrochoid:}
In its singular limit, $c=0.5$, the hypotrochoid has a correlation energy 
of the form $N^{-s}$ 
with $s=1.5$
according to the theory described below.  The data supports this result
 (see Fig.\ \ref{Ecorr}), whereas $s=2$ for smooth cases $c<0.5$. 
As for the previous curves, 
a crossover is observed from $s=1.5$ for smaller $N$ to $s=2$ for larger 
$N$ when $c$ is close to but smaller 
than $0.5$ (see Fig.\ \ref{Ecorr}).

\end{itemize}

\subsection{Regular polygons}

\subsubsection{Schwarz-Christoffel map}
Another kind of domains we have been interested in are regular polygons. 
The conformal map applying the interior of the unit circle into the interior 
of a given polygon is known as the Schwarz-Christoffel transformation \cite{SC1} 
given by 
\begin{equation}
F(w)=F(w_0)+c\int^w_{w_0}\prod_{j=1}^{n}(1-\zeta/w_j)^{\alpha_j-1}d\zeta
\label{SC:int-int}
\end{equation}
where $w_j$ are the pre-images of the vertices on the unit circle, 
$\alpha_j \pi$ are the interior angles at the vertices and $w_0$ and $c$ are 
complex constant.
A variation of Eq.\ \ref{SC:int-int} is the map applying the interior of the 
unit circle on the exterior of a polygon, given by \cite{SC1}
\begin{equation}
F(w)=F(w_0)+c\int^w_{w_0}\zeta^{-2} 
\prod_{j=1}^{n}(1-\zeta/w_j)^{\alpha_j-1}d\zeta.
\label{SC:int-ext}
\end{equation}
Finally, the map we are interested in which takes the exterior of the unit 
circle into the exterior of the polygon is obtained by replacing $w$ by 
$1/w$ in Eq.\ \ref{SC:int-ext}.

In order to compute these maps, we have used the FORTRAN package SCPACK 
developed by L. N. Trefethen \cite{SC2}, as well as the MATLAB SC Toolbox 
developed by T. A. Driscoll \cite{SC1}.

Let us remark that in order to fullfil the condition $F(w) \to w$ as 
$w \to \infty$, we need to choose the size of our regular polygons properly.



\subsubsection{Symmetry breaking}
As for the other curves considered, we observe a symmetry breaking 
of the placement of charges on regular polygons. 
However, the process is slightly different from the one 
observed for the hypotrochoid or the ellipse. Indeed, for both these curves, 
in the singular limit, the symmetry is broken for any finite value of $N$. 
Regular polygons do not present cusps, but corners. Depending on the 
value of the interior angles, the symmetry breaks only for $N$ larger than a 
threshold value $N^*$. For small enough 
values of $N$, the charges remain far enough from the corners to be unaffected
by the charges sitting on the other sides of the polygon. When $N$ 
increases, the charges closest to the corner slightly shift to avoid being 
just in front of the charges on the neighboring side. 
This effect is much more pronounced on curves presenting cusps, because the 
distance between charges placed on two neighboring branches decreases much 
faster as one approaches the cusp. Therefore, the symmetry breaks for any 
finite $N$.

\begin{figure}[htbp]
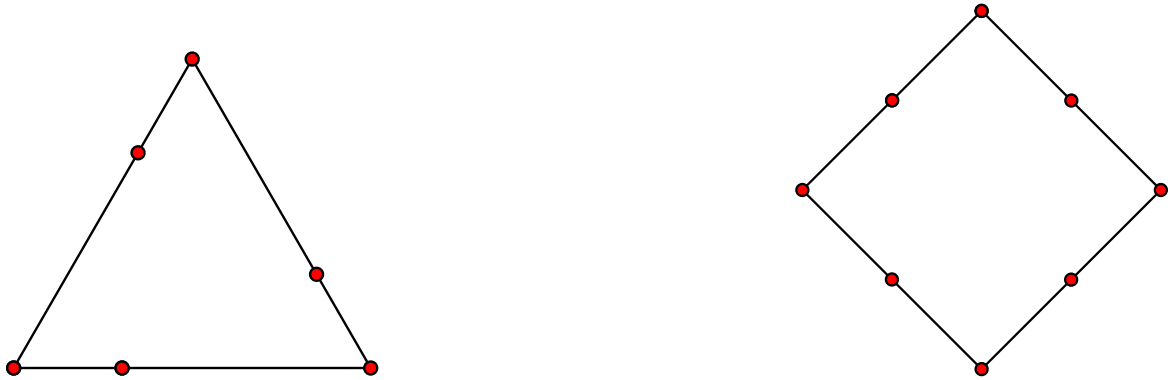

\centering
\includegraphics[width=5cm]{trian-equi-6.eps}
\hfill
\includegraphics[width=5cm]{square-8.eps}
\caption{Placement of 6 charges on an equilateral triangle and of 8 charges 
on a square; the symmetry is broken in the first case and preserved in 
the second.}
\label{broken-unbroken}
\end{figure}

Figure \ref{broken-unbroken} shows the placement of 6 charges on an 
equilateral triangle and of 8 charges on a square. Due to the small value 
of the interior angles in the case of the triangle, the symmetry is  
broken, while it is preserved for the square.

\subsubsection{Minimum energy configuration}
\begin{figure}[htbp]
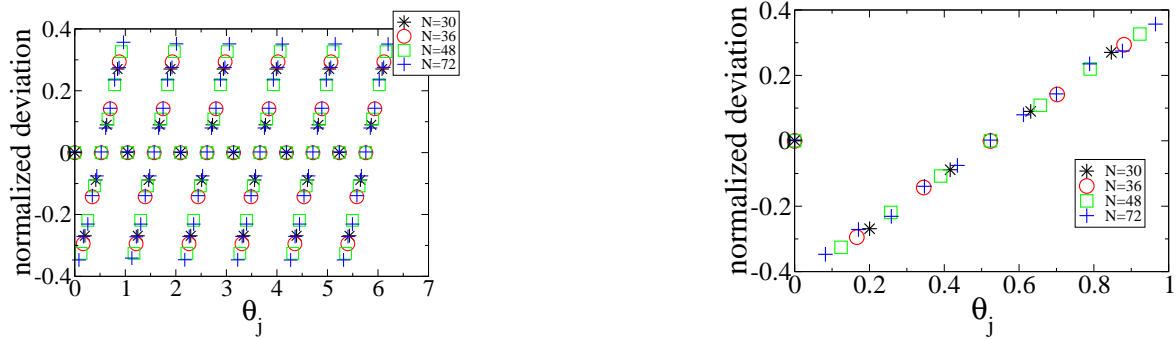

\centering
\includegraphics[width=6cm]{phi-hexa-1.eps}
\hfill
\includegraphics[width=6cm]{phi-hexa-1bis.eps}
\caption{Deviation function $\psi$ for the regular hexagon}
\label{psi-hexa}
\end{figure}

In order to characterize the minimum energy configuration, 
we have computed the 
deviation function $\psi(\theta)$ for the regular hexagon, 
for different values of $N$. 
The results are presented in Fig.\ \ref{psi-hexa}. The charges sitting at the 
corners remain unshifted, while the ones sitting on one side of the hexagon 
shift toward the closest corner. If the number of charges on one side is odd 
(like for $N=30$, $N=36$, $N=48$ and $N=72$), the central charge does 
not shift.  

\begin{figure}[htbp]
\centering
\includegraphics[width=8cm]{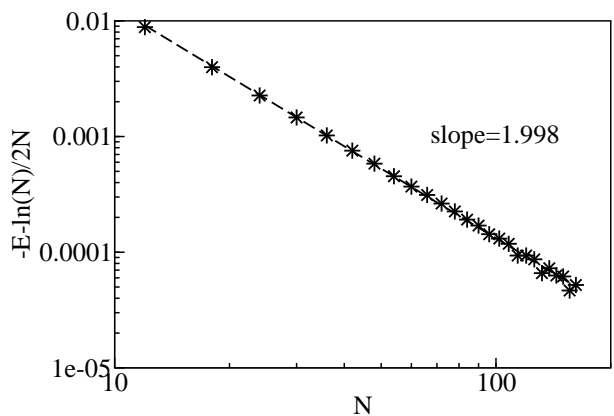}
\caption{Dependence of the correlation energy on $N$ for the regular hexagon.}
\label{hexa-E}
\end{figure}

\begin{figure}[htbp]
\centering
\includegraphics[width=8cm]{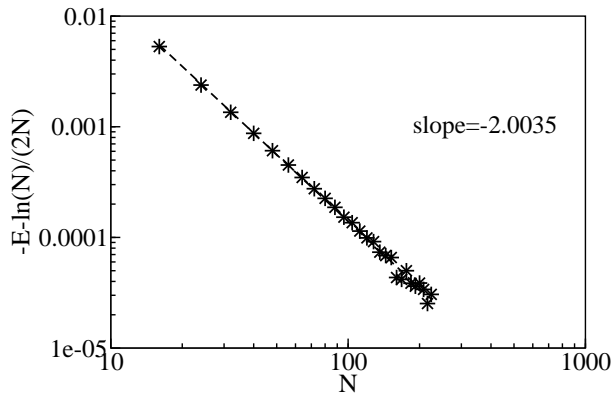}
\caption{Dependence of the correlation energy on $N$ for the regular octagon.}
\label{octo-E}
\end{figure}

For regular polygons, the correlation energy depends on $N$ as 
$N^{-s}$ with $s=2$
according to the theory described below. 
Figures \ref{hexa-E} and \ref{octo-E} 
show this 
dependence for a regular hexagon and a regular octagon. 

\subsection{Scaling arguments for the energy dependence}

The different dependences of the energy on the number of charges $N$
can be easily understood.

The energy itself is of order unity for a two-dimensional conductor of
arbitrary size.  There are $N^2$ interactions, each of strength
$1/N^2$.  The size is picked so that this leading order term in the
correlation energy is zero for when the charges are in their unshifted
position. The shift in position for each particle is of order $1/N$
and that all the particles and interactions participate roughly
equally in the shift.  Since the energy obeys a variational principle,
the energy varies as the square of the shift, or as $1/N^2$.

\begin{figure}[htbp]
\centerline{\includegraphics[width=6cm]{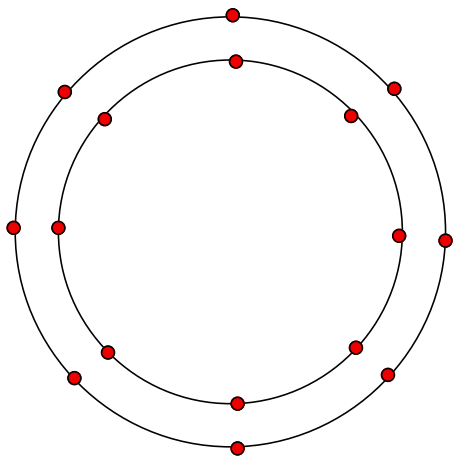}}
\caption{Two concentric circles, with the charges lined up.}
\label{maxen}
\end{figure}

\begin{figure}[htbp]
\centerline{\includegraphics[width=6cm]{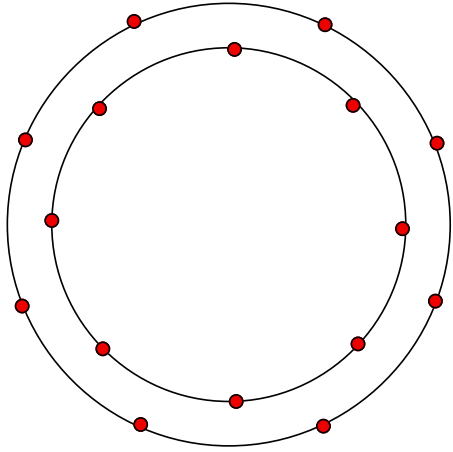}}
\caption{Two concentric circles, shifted.}
\label{minen}
\end{figure}

To find the shift in the singular case, consider first a simplified
model in which charges appear in two concentric closely
spaced circles. To make all the important distances comparable take
the spacing between the circles to be the same as the spacing
between the charges in each circle, i.e. $2\pi/N$.  Compare two
cases, one with the charges lined up (see Fig.\ \ref{maxen}); 
the other with the inner
circle rotated so that no two charges are very close to one
another (see Fig.\ \ref{minen}). 
In the second case, the distance between the closest
charges has increased roughly by a factor of two.  These closest
interactions, each of order $1/N^2$ have changed by an additive
term of order $(1/N^2) \ln 2$.  Thus the entire change in
correlation energy is of order $1/N$.  

The very same argument applies to the hypotrochoid, except
in that it applies to only some of the charges. Look at the
small distance $s$ from the point of the hypotrochoid.  There
are, once gain, two almost parallel lines. At a distance $s$
we have a number of charges, $J$, given by $(J/N)^2\sim s$ and a
separation between the two lines given by $\delta \sim
(j/N)^3 $.  The separation between two changes on a single
branch is the J derivative of $(J/N)^2$ or $\sim J/N^2$. 
The number of charges sensitive to the distortion effects
are the ones with $\delta$ less than or of the order of
this separation.  Hence we have
$$
(J/N)^3  \sim (J/N) (1/N)
$$
or a number of particle $J\sim N^{1/2} $ whose interaction
with the nearest neighboring charges can be changed by an amount
of order unity.  These particles each have an interaction
of order $1/N^2$, so that the total change is
$1/N^{3/2}$, as observed numerically.

For the very skinny ellipse, or for the line segment, all
the particles participate, as happens with the concentric circles, so that
the change is of order $1/N$.

For a polygon, only the very few particles nearest the
corner participate.  Hence the singular change in the
energy is of order $1/N^2$.         

\begin{figure}[htbp]
\centerline{\includegraphics[width=10cm]{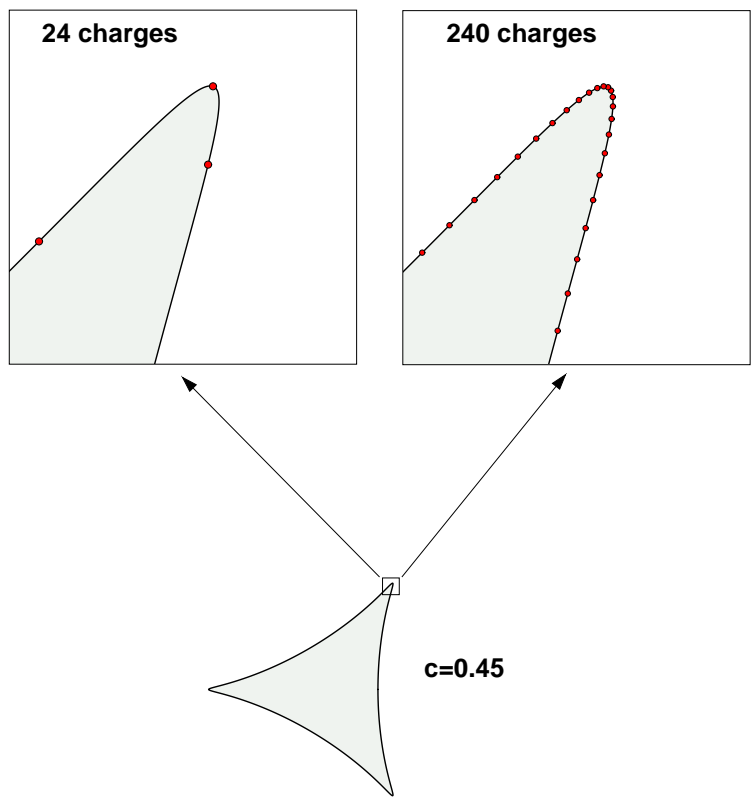}}
\caption{Charge placement for the hypotrochoid with $c=0.45$, for 24 and 
240 charges.}
\label{zoom}
\end{figure}

The crossover observed for values of $c$ close to but smaller than $c^*$ can 
be understood as follows. For a sufficiently small number of charges, the 
charges behave as if the curve were singular: the symmetry is broken and the 
above arguments apply. But when $N$ increases, the smoothness of the curve 
prevails, the symmetry is restored and the correlation energy goes as 
$N^{-s}$ with $s=2$. This is illustrated in Fig.\ \ref{zoom}, where the 
placement 
of charges near a cusp is shown for the hypotrochoid with $c=0.45$, for $N=24$ 
and $N=240$.

\subsection{Symmetry breaking} 
We can also estimate the energy of the symmetric placement of the
charges (See Appendix \ref{sec:apndxb}). By considering a test
(possibly non-optimal) {\em uniform symmetric configuration} $w_j =
\exp(2 \pi i j/N + i \theta_0)$, we see that, with appropriate choices
of $\theta_0$, the minimum energy symmetric configuration for an
ellipse has an energy
$$
E_n \simeq -\frac{\log N}{2N} - \begin{cases}   \frac{1}{2}
\left[N^{-1} - N^{-2}\right]\log(1 - e^{- N \epsilon})  
&  N \text{ odd},  \\
 N^{-2}\log(1 - e^{- N \epsilon/2}) + N^{-1}
\log(1 - e^{- N \epsilon}) & N \text{ even},
\end{cases}
$$
For the hypotrochoid, we have
$$
E_n \simeq -\frac{\log N}{2N} - \begin{cases}C N^{-2} e^{- N \epsilon} & N \neq 0 \bmod 3, \\
C' N^{-2} e^{- N \epsilon/3} + C'' N^{-2} e^{- N \epsilon} & N = 0 
\bmod 3,
\end{cases}
$$ where we expect that, for an appropriate $\theta_0$, the constants
$C,C',C''$ are all positive (See Appendix~\ref{sec:apndxb}). For $N \gg
1$ and $\epsilon > 0$,
$$
-\log(1 - e^{- N \epsilon/2}) \gg -\log(1 - e^{- N \epsilon}), \quad \quad
e^{- N \epsilon/3} \gg  e^{- N \epsilon}.
$$

Consequently, if $N$ is odd, rather than distributing the charges
uniformly on the ellipse, a ``lower'' energy symmetric state would
correspond to having $(N-1)/2$ charges on one ``branch'' of the
ellipse (say between $\theta = 0$ and $\pi$ on the unit circle, and
having $(N+1)/2$ charges on the other branch, with positions
corresponding to uniform distribution of $N-1$ and $N+1$ charges
respectively. For large $N$, the energy of such a ``symmetric'' state
with charges at the sharp ``ends'' of the ellipse will be
$$ 
E_n \simeq -\frac{\log N}{2N} + D_1 N^{-2}\log(1 - e^{- N \epsilon/2}) -
D_2 N^{-1} \log(1 - e^{- N \epsilon}),
$$ where $D_1$ and $D_2 > 0$ are constants that depend on whether $N$
is even or odd. A similar argument for the hypotrochoid yields
$$ 
E_n \simeq -\frac{\log N}{2N} - C N^{-2} e^{- N \epsilon/3},
$$ with a positive constant $C$ that depends on whether or not $N$ is
divisible by 3.

The above arguments show that, the ``symmetry broken'' states have
$$
E_n \simeq -\frac{\log N}{2N} - \begin{cases} A N^{-1} & \text{ellipse}
 \\  B N^{-3/2} & \text{hypotrochoid} \end{cases}
$$ with $A,B > 0$. Comparing the two expressions, we see that the
crossover in the energy scaling, as well as the symmetry breaking
transition occur at
$$
\epsilon \sim \begin{cases} N^{-1}(1-e^{-A/D_2}) & \text{ellipse,} \\
3 N^{-1}( \frac{1}{2}\,{\log N} + F) & \text{hypotrochoid,} \end{cases}
$$ where $F = \log B - \log C$.

\begin{figure}
\centering
\includegraphics[width=\hsize]{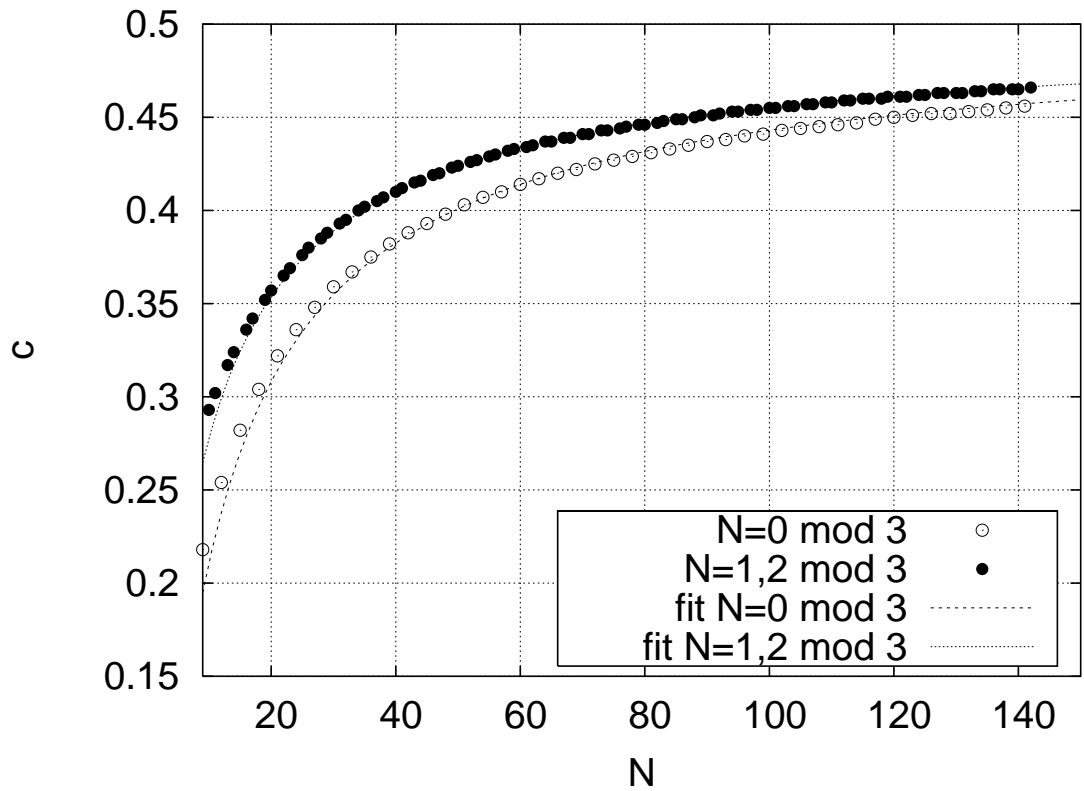}
\caption{A comparison between the predicted and the numerically observed symmetry breaking.}
\label{fig:symm_break}
\end{figure}

We compare the formula for the
hypotrochoid with numerical observations. For the hypotrochoid, the
above formula yields
$$
c = \frac{1}{2} \exp\left[ -\frac{3}{2} \frac{\log N}{N} + \frac{3 F}{N}
\right].
$$ Fig.~\ref{fig:symm_break} is a plot of the $c$--value for the
symmetry breaking, as a function of $N$, for the hypotrochoid. There
is one fitting parameter, $F$, which can depend on whether or not $N$
is divisible by 3.We see that the theory agrees very well with
the numerics, both for $N = 0 \bmod 3$ and $N \neq 0 \bmod 3$.
When $ N=0 mod 3 $ there are three "corners"  which show a distortion from the symmetry breaking while in the other cases there is but one.  Hence the former case has the most to gain from the symmetry breaking, and so we expect the asymmetry to persist down to lower values of $c$ for that case.  The data supports this expectation.     

\subsection{Universality in the Charge locations for singular curves}

We first consider the problem of the placement of $N$ identical charges of 
strength $1/N$ on the line segment $\Gamma_1=[-1,1]$. 
The Fekete polynomial $f_N(x)$ for $\Gamma_1$ is given by 

$$f_N(x)=\prod_{j}^N |x-\zeta_j|$$ 

where $\zeta_j$ are defined by 
$$
\prod_{i=1}^N\prod_{j \neq i}^N |\zeta_i -\zeta_j| 
=\max_{x_1,\ldots,x_N \in [-1,1]} \prod_{i=1}^N\prod_{j \neq i}^N |x_i - x_j|.
$$ 
 
$f_N$ obeys the equation \cite{stieltjes}

$$(1-x^2)f_N''+N(N-1)f_N=0.$$

\begin{figure}[htbp]
\vskip 1cm
\centerline{\includegraphics[width=6cm]{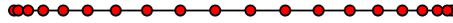}}
\caption{Placement of 20 charges on a line segment.}
\label{line}
\end{figure}

The solution can be expressed as
$$
f_N(x)=c_N (x^2-1) P'_{N-1}(x)=2^N 
{\binom{2N}{N}}^{-1}\{P_N(x)-P_{N-2}(x)\},
$$ 
where $P_k$ denotes the Legendre polynomial of degree $k$
\cite{stieltjes,szego_polynomials,stieltjes_review}. $N$ identical
charges therefore place themselves at the ends of the segment and at
the positions of the extrema of the Legendre polynomial of degree
$N-1$, or equivalently at the intersections of the Legendre
polynomials of degrees $N$ and $N-2$.  Figure \ref{line} represents
the case $N=20$. The charges accumulate toward the ends. This is
consistent with the solution of the continuum problem, i.e., the
equilibrium charge density on the line segment $\Gamma_1=[-1,1]$,
corresponding to the situation $N \rightarrow \infty$, and given by
$$\rho(x)=\frac{1}{\pi} \frac{1}{\sqrt{1-x^2}}.$$

One way to characterize the placement of $N$ identical charges on a
line segment is to compute the ratio of two successive intervals
between the charges, $\Delta_{i+1}/\Delta_{i}$ where
$\Delta_i=|x_{i+1}-x_i|$, $i=1$ corresponding to the charge placed at
one extremity of the segment.  The asymptotic dependence of
$\Delta_{i+1}/\Delta_{i}$ on $N$ is given by
$$\Delta_{i+1}/\Delta_{i}=R_i-\frac{A_i}{N^2}+\mathcal{O}(N^4),$$
where $R_i$ and $A_i$ are two constants depending on $i$.

\begin{figure}[htbp]
\centerline{\includegraphics[width=8cm]{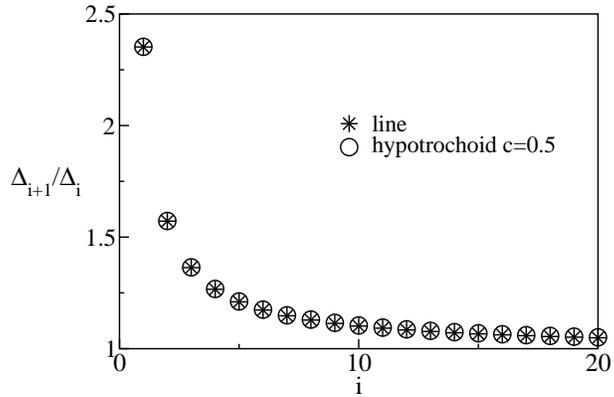}}
\caption{Dependence of $\Delta_{i+1}/\Delta_{i}$ on $i$.}
\label{spacing}
\end{figure}

The star symbols in figure \ref{spacing} show the dependence of 
$\Delta_{i+1}/\Delta_{i}$ on $i$ for the line, obtained for $N=200$. 
As expected, $\Delta_{i+1}/\Delta_{i}$ tends 
to 1 when $i$ increases, i.e., in the central part of the line segment. 

The other singular curve we are interested in is the hypotrochoid with
$c=0.5$. In this singular limit, the charge symmetry is broken for any
finite number of charges $N$. Moreover, in the limit of large $N$, the
placement of the charges near the cusps is similar to the placement of
charges on the line segment. We have projected the charges positions
near one cusp on the symmetry axis of this cusp, and we have computed
the quantities $\Delta_{i+1}/\Delta_{i}$ as defined for the line
segment. The results are shown in Fig.\ \ref{spacing} by circles,
computed for $N=800$. The correspondence with the results for the line
is excellent. This is a signature of universal behavior of the charge
locations close to the cusp {\em even in the symmetry broken regime}.

\begin{figure}[h]
\centering
\includegraphics[width=5cm]{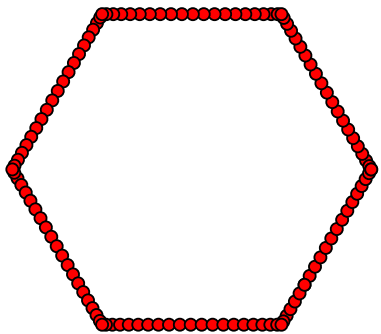}
\caption{Placement of 120 charges on a regular hexagon.}
\label{breaking2}
\end{figure}

\begin{figure}[htbp]
\centering
\includegraphics[width=5cm]{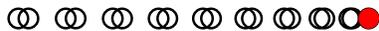}
\caption{Projection on the horizontal axis of the charge positions 
on two opposite sides of the regular hexagon of Fig. \ref{breaking2}; 
the filled circle is the corner charge.}
\label{breaking3}
\end{figure}

Let us remark that the situation is completely different in the case of 
regular polygons, where we consider the placement of charges near a corner 
instead of a cusp. This 
is illustrated in Figs.\ \ref{breaking2} and \ref{breaking3}, 
which show the placement of 120 charges on a regular hexagon. 
The second figure shows the projection of 
the charges on the horizontal axis. The shift between two facing charges is 
very small, and their projections remain paired. This behavior is 
completely different from the one observed for the line and the hypotrochoid

\vspace{1.5cm}
\section{Discussion}

We have investigated some aspects of the electrostatics of discrete
charges in two dimensions. The problem is of significant interest
because of its connections to approximation theory
\cite{fekete_review, deift}, constructing conformal maps
\cite{Pom_book} and Integrable systems
\cite{Wiegmann00,tau_function}. 

A dynamical version of this problem is also closely related to various
Laplacian growth models because of the deep connections between
conformal maps and Laplacian growth \cite{Hastings_LG,
FPD_LG,Wiegmann_LG}. The dynamics of discrete charges in two
dimensions, when they are confined by appropriate potentials \cite{HKD02}, is
therefore relevant to electronic droplet in Quantum-Hall systems
\cite{Wiegmann_LG}, and to the Laplacian growth problems including
Hele-Shaw \cite{FPD_LG} and DLA \cite{Hastings_LG}. The rich structure
of the static problem certainly leads one to expect a similar richness
in the associated dynamics, and we will investigate this question
further.

In addition, the static problem is also a prototype for the following
class of problems --

{\em 
Find the ``optimal'' way to discretize a continuum quantity.}

As stated, this discrete optimization formulation is much too general,
for us to say anything useful about it. Given a specific problem
however, there is an appropriate continuum quantity, and an
appropriate notion of optimality. Such problems arise in many places
including approximation theory \cite{approx,approx2}, optimal
transport \cite{mccann}, in discretizing integrable systems
\cite{discrete1,discrete2}, {\em etc}.

As in the case of 2-D electrostatics, these discrete problems can have
a much richer structure than the corresponding continuum problem. In
particular, discretizing the continuum solution in an obvious manner
{\em will not}, in general, be a good solution for the discrete
problem. One explicitly needs to account for the discreteness in the
formulation. As we show in Sec.~1, the locations of the discrete
charges correspond to discretizing a density $\rho$ which is {\em not}
the continuum solution that minimizes the electrostatic
energy. Rather, the density $\rho$ minimizes the {\em modified}
energy, $E_{s} + E_{corr}$, which are given in Eqs.~(\ref{E_svar}) and
(\ref{E_corr}).

We can increase the ``discreteness'' of the problem, by making curve
more singular, or equivalently reducing the total number of charges
$N$, for a fixed curve. As we increase the ``discreteness'', we get
further away from the continuum solution. In particular, we have
interesting, qualitatively new phenomena including {\em symmetry
breaking}, which implies, among other things, the lack of a unique
minimizing configuration. This is in sharp contrast to the continuum
problem, which is {\em convex}, and therefore has a unique
solution. Discreteness can thus introduce non-convexity into a problem
whose continuum version is convex, and thereby change the problem
qualitatively \cite{convex1,convex2}.

One of the results from this paper is the universal behavior of the the
charge distribution , {\em viz}., the $\psi$ function in the
neighborhood of cusps in the symmetric regime. This behavior comes out
of minimizing the modified energy $E_s + E_{corr}$. This is no longer
valid in the symmetry broken regime, as illustrated by
Figs.~\ref{sk-ell}, \ref{sk-hypo} and \ref{fig:phi_function}. Despite
this, universal behavior persists even in the symmetry broken regime,
as we illustrate in Fig.~\ref{spacing}. It is interesting to
understand the nature of this universality, and we will explore this
further in the future.


\section*{Acknowledgments}
We would like to thank Paul Wiegmann, Thomas Erber, David Nelson, and
Michael Brenner for helpful conversations.  This research was supported
by the University of Chicago MRSEC under NSF grant DMR-0213745. Additional
support came from the NSF-DMR under grant 094569.

\appendix

\section{Saddle point evaluation of $s_m^{(i)}$} \label{sec:apndxa}

 To evaluate the successive terms $s_m^{(i)}$ in the series solution
Eq.~(\ref{eq:series_soln}), we need the following result --
\begin{lemma}
Let $\alpha$ be a positive real number. Then,
$$
\sum_{n = \left[\frac{m+1}{2}\right]}^{2 m}
\left(\frac{m}{n+m}\right) \binom{m+n}{2m - n} \alpha^n = \frac{1}{3}\beta^m \left( 1 +
O\left(\frac{1}{m}\right) \right),
$$
where
$$ \beta = \frac{(\alpha + \sqrt{4 \alpha +
\alpha^2})^3}{8 \alpha}
$$
and
$$
\binom{m+n}{2m-n} = \frac{(m+n)!}{(2m-n)!(2n-m)!}
$$
denotes the binomial coefficient.
\label{lem:saddle}
\end{lemma}

\begin{proof}
Setting $n = my$, replacing the sum over $n$ by an integral with
measure $m dy$, using Stirling's formula
$$ p! = \sqrt{2 \pi p} \left(\frac{p}{e}\right)^p\left[1 +
O\left(\frac{1}{p}\right)\right],
$$
we see that the sum
$$
\sum_{n = \left[\frac{m+1}{2}\right]}^{2 m} \left(\frac{m}{n+m}\right)
\left(\begin{array}{c} m+n \\ 2m-n \end{array}\right) \alpha^n
$$
reduces to
$$
\frac{m}{\sqrt{2 \pi m}}
\int_{y = 1/2}^{2} \frac{\alpha^{m y}}{1+y}
\sqrt{\frac{1+y}{(2 - y)(2y-1)}} \left[\frac{(1+y)^{(1+y)}}{(2 -
y)^{(2-y)} (2y-1)^{(2y-1)}}\right]^m dy
$$
with an $O\left(\frac{1}{m}\right)$ relative error.

We will evaluate this integral by the saddle point method. Let $g(y)$
be defined as
$$
g(y) = \log\left[\frac{\alpha^y (1+y)^{(1+y)}}{(2-y)^{(2-y)}(2y-1)^{(2y-1)}}\right],
$$
so that the integral reduces to
$$
I = \frac{m}{\sqrt{2 \pi m}}
\int_{y = 1/2}^{2} \frac{e^{m g(y)}}{\sqrt{(1+y)(2 - y)(2y-1)}} dy.
$$
The maximum for $g(y)$ is at $y^*$ given by
$$
\log(\alpha) + \log(1+y^*) + \log(2-y^*) - 2 \log(2 y^* - 1) = 0.
$$
This yields the equation
$$
\frac{\alpha(1+y^*)(2-y^*)}{(2 y^* - 1)^2} = 1.
$$
Solving this quadratic equation, the relevant solution is
$$
y^* = \frac{1}{2} + \frac{3}{2} \sqrt{\frac{\alpha}{4+\alpha}},
$$ For $0 < \alpha < \infty$, we get $y^* \in (1/2,2)$, so that the
extremum lies within the range of integration.
$$
\exp(g(y^*)) = \frac{1}{\alpha}\left(\frac{2 y^* - 1}{2 -
y^*}\right)^3 = \frac{1}{\alpha} \left(\frac{2
\sqrt{\alpha}}{\sqrt{4+\alpha} - \sqrt{\alpha}}\right)^3 = \frac{(\alpha + \sqrt{4 \alpha +
\alpha^2})^3}{8 \alpha} = \beta.
$$
Expanding $g(y)$ about $y^*$, we get
$$
g(y) = g(y^*) - \frac{9}{2 (1+y^*)(2-y^*)(1-2y^*)}(y-y^*)^2 + O((y-y^*)^3)
$$ so that the point $y^*$ is a maximum for $g$. We are now in a
position to apply the saddle point method, and for large $m$ we obtain
$$
I \sim \frac{1}{3}\beta^m
$$
asymptotically, with $O(1/m)$ corrections.
\end{proof}

\section{The energy of symmetric configurations for singular shapes}
\label{sec:apndxb}

We estimate the energy of the symmetric configuration obtained by
distributing the $N$ charges uniformly on the unit circle, {\em i.e},
the points on the curve are $z_j = F(w_j) = F[e^{i (\theta_0 + 2 \pi
j/N)}]$. We are especially interested in situations close to the
``singular'' shape ($c = 1$ for the ellipse and $c = 1/2$ for the
hypotrochoid). For this choice of $w_j$, we have
$$
s_k = \sum_{j = 0}^{N-1} w_j^k = \begin{cases} 0 & k \neq  0 \bmod N
\\ N e^{i k \theta_0} & k = 0 \bmod N \end{cases}
$$

As we show in Sec.~\ref{sec:real_space}, the deviation of the energy from
that of a unit circle can be written as 
$$
\Delta E_N = \Re \int \frac{dw}{4 \pi i w} \frac{w^N}{w^N-1} \ln
w F^\prime(w).
$$
Writing this expression in terms of $s_k$ and the matrices $a_{kl}$
 and $c_k$, we get
$$ \Delta E_N = \frac{1}{2} \sum_{k \geq 1} \sum_{l \geq 1} a_{kN,lN}
e^{-i(k+l)N \theta_0}- \frac{1}{2N}
\sum_{k \geq 1} c_{kN} e^{- i k N \theta_0}.
$$

We will first evaluate this sum for the ellipse.  Using the
expressions for $a_{kl}$ and $c_k$ in Eqs.~(\ref{a:ellipse}) and
(\ref{c:ellipse}) respectively, we get
\begin{equation}
\Delta E_N^{ell} = \frac{1}{2} \sum_{q \geq 1} \frac{c^q}{q} \delta_{q,kN}
e^{-2ikN \theta_0}- \frac{1}{2N}
\sum_{q \geq 1} \frac{c^q}{q}\delta_{2q,kN} e^{- i k N \theta_0}
\label{eq:energy_ellipse}
\end{equation}
We need to do the cases $N$ odd and $N$ even separately.

If $N$ is odd, $2q = kN$ implies that $q$ is a multiple of $N$ and $k$
is even. Therefore $2q = kN$ implies $q = jN$ and $k =
2j$. Consequently, Eq.~(\ref{eq:energy_ellipse}) yields
$$
\Delta E_N^{ell} = \frac{1}{2N} \sum_{k \geq 1}\frac{[c e^{-2 i \theta_0}]
^{Nk}}{k} - \frac{1}{2 N^2} \sum_{j \geq 1} \frac{[c e^{-2i\theta_0}]^{Nj}}{j}.
$$
If $N$ is even, $2q = k N$ implies that $q$ is a multiple of $N/2$ and
there are no restrictions on $k$. Therefore $2q = kN$ implies $q =
k(N/2)$, and $k \geq 1$ is any natural
number. Eq~(\ref{eq:energy_ellipse}) now yields
$$
\Delta E_N^{ell} = \frac{1}{2N} \sum_{k \geq 1}\frac{[c e^{-2 i \theta_0}]
^{Nk}}{k} - \frac{1}{N^2} \sum_{k \geq 1} \frac{[c e^{-2i\theta_0}]^{Nk/2}}{k}.
$$
We can evaluate these sums using the identity
$$
\sum_{k = 1}^{\infty} \frac{e^{-\lambda k}}{k} = -\log(1-e^{-\lambda}).
$$
This yields the following {\em exact} result for the ellipse
\begin{equation}
\Delta E_N^{ell} = \begin{cases}
 \left[\frac{1}{2 N^2} - \frac{1}{2 N}\right] \log(1 - e^{- N (\epsilon + 2 i \theta_0)}) & N \text{ odd } \\
 - \frac{1}{2 N}\log(1 - e^{- N (\epsilon + 2 i \theta_0)}) +
\frac{1}{N^2} \log(1 - e^{- N (\epsilon + 2 i \theta_0)/2})& N \text{ even },
\end{cases}
\label{eq:shift_ellipse}
\end{equation}
where, as defined earlier, $\epsilon = \log(c) \approx 1-c$

\subsection{Hypotrochoid}

We can do a similar calculation for the hypotrochoid. Using the
expressions in Eqs.~(\ref{a:hypotrochoid}) and (\ref{c:hypotrochoid}),
we obtain
\begin{equation}
\Delta E_N^{hypo} =  \frac{1}{2} \sum_{q \geq 1} \frac{c^q}{q} \left[\sum_{p = 0}^q \delta_{q+p,kN} \delta_{2q-p,lN} \binom{q}{p} - \frac{1}{N}2^q\delta_{3q,jN} \right]e^{-3iq \theta_0}
\label{eq:energy_hypotrochoid}
\end{equation}
We will now look which values of $q$ and $p$ contribute in the
above summation.
$$
q+p = 0, 2q - p = 0 \bmod N \quad \implies \quad 3q = 0, 3 p = 0 \bmod N.
$$
We therefore need to do the cases $N = 0 \bmod 3$ and $N \neq 0 \bmod
3$ separately. If $N \neq 0 \bmod 3$, $3q = 0 \bmod N$ and $3p = 0
\bmod N$ imply that that $q = j N$, $p = m N$, where $j,m$ are integers.
In this case, Eq.~(\ref{eq:energy_hypotrochoid}) yields
\begin{align}
\Delta E_N^{hypo} & = \frac{1}{2N} \sum_{j \geq 1} \frac{[c e^{-3 i \theta_0}]
^{jN}}{j} \left[\sum_{m = 0}^j \binom{jN}{mN} - \frac{1}{N}2^{jN}
\right] \nonumber \\
& = I + II
\label{eq:hypo_nondiv}
\end{align}
We can estimate part $II$ of the sum as above. We have
$$
- \frac{1}{2N^2} \sum_{j \geq 1} \frac{[c e^{-3 i \theta_0}] ^{jN}}{j}
2^{jN} = \frac{1}{2N^2} \log\left[ 1- [2 c e^{-3 i
\theta_0}]^{N}\right]
$$
To estimate part $I$, we use the identity
$$
\sum_{\beta = 0}^{N-1} \left( 1 + e^{2 \pi i \beta/N}\right)^{jN}
= N \sum_{m = 0}^j \binom{jN}{mN}.
$$
Using this, and the above considerations, we get
$$
I = \frac{1}{2N^2} \sum_{j \geq 1} \frac{[c e^{-3 i \theta_0}]
^{jN}}{j} \sum_{\beta = 0}^{N-1} \left( 1 + e^{2 \pi i \beta/N}\right)^{jN}
$$
Interchanging the order of the summations, we get
$$
I = \frac{1}{2N^2} \sum_{\beta = 0}^{N-1} \log\left[ 1 - \left[c e^{-3 i \theta_0}\left( 1 + e^{2 \pi i \beta/N}\right)\right]^N \right]
$$
Since
$$
\left( 1 + e^{2 \pi i \beta/N}\right)^N = 2^N \cos^{N}\left(\frac{\pi \beta}{N}\right) e^{i \pi \beta},
$$
writing $\sum_{\beta = 0}^{N-1} = \sum_{\beta = -N/2}^{N/2-1}$, and
using $2c = e^{- \epsilon}$ we get
\begin{align*}
I & =  \frac{1}{4N^2} \sum_{\beta = N/2}^{N/2-1} \log\left[ 1 + e^{-2 N \epsilon} \cos^{2N}\left(\frac{\pi \beta}{N}\right) - 2 (-1)^{\beta}e^{-N \epsilon}\cos^{N}\left(\frac{\pi \beta}{N}\right) \cos(3 N \theta_0)\right]
\end{align*}
We will estimate this term for $\theta_0 = 0$. We get,
$$
I = \frac{1}{2N^2}  \sum_{\beta = N/2}^{N/2-1} \log\left[ 1 - (-1)^\beta e^{-N \epsilon} \cos^{N}\left(\frac{\pi \beta}{N}\right)\right].
$$
We will estimate the sums over odd $\beta$ and even $\beta$
separately. We get
$$
I = \frac{1}{2N^2}  \left[\sum_{\beta \mbox{ odd}} \log\left[ 1 + e^{-N \epsilon} \cos^{N}\left(\frac{\pi \beta}{N}\right)\right] + \sum_{\beta \mbox{ even}} \log\left[ 1 - e^{-N \epsilon} \cos^{N}\left(\frac{\pi \beta}{N}\right)\right]\right]
$$
Note that the terms with odd $\beta$ are all positive, and the terms
with even $\beta$ are all negative.


It is easy to verify that, for an appropriate constant $C$, we have
$$ e^{- \theta^2/2}(1 - C \theta^4) \leq \cos(\theta) \leq e^{- \theta^2/2},\quad \quad  -\pi/2 \leq \theta \leq \pi/2.
$$
Using this, for $-N/2 \leq \beta \leq N/2$, we get
$$
\exp\left(- \frac{\pi^2 \beta^2}{2N}\right) \left(1 - \frac{C' \beta^4}{N^4}\right)^N \leq \cos^{N}\left(\frac{\pi \beta}{N}\right) \leq \exp\left(- \frac{\pi^2 \beta^2}{2N}\right),
$$
where $C'$ is an appropriate constant. It is clear from the above
inequalities, that $\cos^{N}\left(\frac{\pi \beta}{N}\right)$ is
significantly different from zero only for $|\beta| \lesssim \sqrt{N}
\ll N/2$ as $N \rightarrow \infty$. Also, if $\beta \sim O(\sqrt{N})$ we have
$$
\left|1 - \left(1 - \frac{C' \beta^4}{N^4}\right)^N\right| \lesssim O\left(\frac{1}{N}\right)
$$
in the region where the exponential $e^{- \pi^2 \beta^2/2N}$ is not
exponentially small. Consequently, the relative error in the
approximation
$$
\cos^{N}\left(\frac{\pi \beta}{N}\right) \approx \exp\left(- \frac{\pi^2 \beta^2}{2N}\right),
$$
goes to zero as $N \rightarrow \infty$. We thus obtain
\begin{align*}
\frac{1}{4N^2} \sum_{\beta = -N/2}^{N/2-1} \log\left[ 1 - \alpha \cos^{N}\left(\frac{\pi \beta}{N}\right) \right] & \approx \frac{1}{4N^2} \int_{-N/2}^{N/2} \log\left[ 1 - \alpha \exp\left(\frac{-\pi^2 \beta^2}{2N}\right) \right] d \beta \\
& \approx \frac{1}{4 \sqrt{\pi N^3}}\int_{-\infty}^{\infty}
\log\left[1 - \alpha \exp(-x^2/2)\right] dx,
\end{align*}
where we set $x = \pi \beta/\sqrt{N}$, and the relative error is
$O(N^{-1})$ as $N \rightarrow \infty$.

Note also, if the sum over $\beta$ was restricted to either only odd
or only even $\beta$, we have to multiply the final integral by $1/2$
to obtain the right answer. This can be justified since the integrand
only varies on a scale $\beta \sim \sqrt{N} \gg 1$, so that the sum
over the even $\beta$ should very nearly equal the sum over the odd
$\beta$.

We will now evaluate the integral in the last line for $|\alpha| \leq
1$. Using
$$
\log(1-y) = - \left[y + \frac{y^2}{2} + \frac{y^3}{3} + \cdots\right],
$$
and integrating the resulting series termwise, we get
$$
\int_{-\infty}^{\infty} \log\left[1 - \alpha \exp(-x^2/2)\right] dx = - \sqrt{2 \pi}\sum_{k = 1}^{\infty} \frac{\alpha^k}{k^{3/2}}.
$$
We define
$$
G(\alpha) = \sum_{k = 1}^{\infty} \frac{\alpha^k}{k^{3/2}}, \quad |\alpha| < 1.
$$
It is easy to see that, at $\alpha = 1$, the sum is the Riemann zeta
function $G(1) = \zeta(3/2)$. Also, the series for $G'(\alpha)$
diverges like $C/\sqrt{1 - \alpha}$ as $\alpha \rightarrow
1$. Consequently, for $\alpha$ close to 1, we have
$$
G(\alpha) = \zeta(3/2) - C \sqrt{1 - \alpha} + O(|1-\alpha|^{3/2}).
$$
Also, $G(0) = 0,G'(0) = 1$ and $G''(0) = 1/\sqrt{8}$.

Using the above considerations, we can estimate the term $I$ as
$$
I = - \frac{1}{4 \sqrt{2N^3}}\left[G(e^{-N\epsilon}) + G(-e^{-N
\epsilon})\right] + O\left(\frac{e^{-N\epsilon}}{N^{5/2}}\right).
$$
Therefore, we have,
$$
I \approx \begin{cases} -\frac{\zeta(3/2)}{8 N^{3/2}} & N \epsilon \ll
1, \\ -\frac{e^{-2N\epsilon}}{8 N^{3/2}} +
O\left(\frac{e^{-N\epsilon}}{N^{5/2}}\right) & N \epsilon \gg 1.
\end{cases}
$$

We now consider the case $N = 0 \bmod 3$. Te only terms in the
summations in eq.~(\ref{eq:energy_hypotrochoid}) that contribute arise
from
$$
q+p = 0, 2q - p = 0 \bmod N \quad \implies \quad 3q = 0, 3 p = 0 \bmod N.
$$
Since $N = 0 \bmod 3$, $3q = 0 \bmod N$ and $3p = 0
\bmod N$ imply that that $q = j L$, $p = m L$, where $j,m$ are integers,
and $L = N/3$. Also, $q+p = 0 \bmod N$ implies that $j+m = 0 \bmod 3$,
and this also implies $2p - q = pN - (p+q)M = 0 \bmod N$.

If $j = 3k -r$, where $r \in \{0,1,2\}$, it follows that $m = 3t + r$,
where $k,r$ are integers. Also, $q = jL = k N - r L$ and $p = m L = t
N + r L$. Consequently, eq.~(\ref{eq:energy_hypotrochoid}) yields
\begin{align}
\Delta E_N^{hypo} & = \frac{1}{2}\sum_{r = 0}^2 \sum_{k \geq 1} \frac{[c e^{-3 i \theta_0}]
^{k N - r L}}{k N - r L} \left[\sum_{t = 0}^{k - [r/2]} \binom{kN -
rL}{tN + rL}\right] - \frac{1}{2NL}\sum_{j \geq 1} \frac{[2 c e^{-3 i \theta_0}]
^{jL}}{j} \nonumber \\
& = I + II
\label{eq:hypo_div}
\end{align}
Part $II$ gives
$$
- \frac{1}{2NL} \sum_{j \geq 1} \frac{[2c e^{-3 i \theta_0}] ^{jL}}{j}
 = \frac{3}{2N^2} \log\left[ 1- [2 c e^{-3 i
\theta_0}]^{N/3}\right]
$$
To estimate part $I$, we use the identity
$$
\sum_{\beta = 0}^{N-1} e^{- 2 \pi i \beta r/3} \left( 1 + e^{2 \pi i \beta/N}\right)^{kN - rL}
= N \sum_{t = 0}^{k - [r/2]} \binom{kN-rL}{tN+rL}.
$$
Using this in eq.~(\ref{eq:hypo_div}), we get
\begin{equation}
I = \frac{1}{2N} \sum_{r = 0}^2 e^{- 2 \pi i \beta r/3} \sum_{k \geq
1} \frac{[c e^{-3 i \theta_0}] ^{kN-rL}}{kN-rL} \sum_{\beta = 0}^{N-1} \left(
1 + e^{2 \pi i \beta/N}\right)^{kN-rL}
\label{eq:intermediate}
\end{equation}
For all complex number $|z| < 1$ and $r = 0,1,2$, we have
$$
3 \sum_{k \geq 1} \frac{z^{3 k - r}}{3k - r} =
\begin{cases} - \log(1-z^3), 
& r = 0, \\
- \log(1-z) - \omega^2 \log(1 - \omega z) - \omega
\log(1 - \omega^2 z), & r = 1, \\
- \log(1-z) - \omega \log(1 - \omega z) - \omega^2 \log(1
  - \omega^2 z), & r = 2,
\end{cases}
$$
$$
\sum_{k \geq 1} \frac{z^{3 k - r}}{3k - r} = -\frac{1}{3}\sum_{s = 0}^2
\omega^{2rs} \log(1 - \omega^s z)
$$
where $\omega = e^{-2 \pi i/3}$ is a cube root of unity.

Interchanging the order of the summations in (\ref{eq:intermediate}),
and using the above identities, we get
$$
I = - \frac{1}{2N^2} \sum_{\beta = 0}^{N-1} \sum_{s = 0}^2 \sum_{r =
0}^2 \omega^{(\beta + 2 s)r} \log\left[ 1 - \omega^s \left[c e^{-3 i
\theta_0}\left( 1 + e^{2 \pi i
\beta/N}\right)\right]^N \right]
$$
Note that
$$
\sum_{r =
0}^2 \omega^{(\beta + 2 s)r} = \begin{cases} 0 & \beta + 2 s \neq 0 \bmod 3 \\
3 & s = \beta \bmod 3. \end{cases}
$$
Therefore,
$$
I = - \frac{3}{2N^2} \sum_{\beta = 0}^{N-1} \log\left[ 1 - e^{- 2 \pi
i \beta/3} \left[c e^{-3 i
\theta_0}\left( 1 + e^{2 \pi i
\beta/N}\right)\right]^N \right]
$$
Breaking this sum up into 6 parts using $\beta \bmod 6$, setting
$\theta_0 = 0 $ and noticing the similarity with the $N \neq 0 \bmod
3$ case, we get
$$
I = \frac{1}{4\sqrt{2N^3}}\sum_{r=0}^2\left[G(\omega^r
e^{-N\epsilon/3}) + G(-\omega^r e^{-N\epsilon/3})\right] +
O\left(\frac{e^{-N\epsilon/3}}{N^{5/2}}\right).
$$

\bibliographystyle{abbrv}

\bibliography{fekete}

\end{document}